\documentclass[twocolumn,showpacs,showkeys,preprintnumbers,amsmath,amssymb]{revtex4}

\usepackage{graphicx}
\usepackage{dcolumn}
\usepackage{bm}

\begin{document}


\title{Emission spectra and quantum efficiency of single-photon sources
\\ in the cavity-QED strong-coupling regime}

\author{Guoqiang Cui and M. G. Raymer}
\affiliation{Oregon Center for Optics and Department of Physics,
University of Oregon, Eugene, Oregon 97403, USA}
\email{raymer@uoregon.edu}

\date{\today}

\begin{abstract}
We derive analytical formulas for the forward emission and side
emission spectra of cavity-modified single-photon sources, as well
as the corresponding normal-mode oscillations in the cavity quantum
electrodynamics strong-coupling regime. We investigate the effects
of pure dephasing, treated in the phase-diffusion model based on a
Wiener-Levy process, on the emission spectra and normal-mode
oscillations. We also extend our previous calculation of quantum
efficiency to include the pure dephasing process. All results are
obtained in the Weisskopf-Wigner approximation for an
impulse-excited emitter. We find that the spectra are broadened, the
depths of the normal-mode oscillations are reduced and the quantum
efficiency is decreased in the presence of pure dephasing.
\end{abstract}

\maketitle

\section{Introduction}

Single-photon sources (SPS) have important uses in quantum
communication \cite{Bouwmeester00, Bennett92}, quantum computing
\cite{Knill01}, and metrology \cite{Giovannetti04}. Since the turn
of this century, significant progress has been made in generating
single photons from a microscopic quantum emitter such as an atom or
ion \cite{Kuhn02, McKeever04}, an organic molecule \cite{Brunel99,
Lounis00, Treussart02}, a semiconductor quantum dot (QD) or
nanocrystal \cite{Michler001, Michler002, Santori01, Santori02,
Moreau01, Zwiller02, Pelton02, Vuckovic03}, or a color center in
diamond \cite{Kurtsiefer00, Beveratos01, Gaebel04}. Most of these
SPS are based on spontaneous emission, whose lifetime eventually
limits the emission rate and linewidth of the SPS, and whose
isotropic nature prevents high collection efficiency. A more
promising scheme for producing well-controlled single photons is
cavity quantum electrodynamics (QED). An atom or QD inside a
high-finesse microcavity is prepared in an excited state and is
allowed to spontaneously emit. The emission rate and linewidth can
be considerably altered \cite{Santori01, Santori02} in the
cavity-QED weak-coupling regime. In the bad-cavity limit, the
altered emission rate is obtained from Fermi's golden rule using a
modified density of states to account for the cavity boundary
conditions. In the weak-coupling regime, the atomic excitation is
irreversibly lost to the continuum of all available photon states,
including the side modes (leak modes) and cavity modes. The side
emission and forward emission spectra are single-peaked, either
enhanced or inhibited.

If the cavity volume is sufficiently small, and the finesse is high
enough such that the coherent interaction rate between the quantum
emitter and cavity exceeds the decay rates of the composite system
to be in the cavity-QED strong-coupling regime, the emission process
is reversible and a photon emitted into the cavity can be coherently
reabsorbed before it is emitted out of the cavity. The initially
excited quantum emitter undergoes single-quantum Rabi oscillations.
The emissions, still allowed to both the side and forward direction
of the cavity, will show double-peaked spectra. In this regime, the
coupling of the emission to the single cavity mode, however, is far
stronger than its coupling to the side modes. If, in addition, there
is almost no dephasing of the quantum emitter during the emission
process, the emission process can be nearly deterministic
\cite{Kuhn02, McKeever04}, emitting a photon into a well-defined
``forward'' beam outside the cavity.

Most if not all single-quantum systems, however, inevitably interact
with certain heat baths, leading to dephasing or loss of coherence,
which results from a randomization of the phases of the emitter's
wave functions by thermal fluctuations in the environmental fields.
Population relaxation processes contribute to dephasing with a
dephasing rate given by half the population decay rate. It is often
necessary to account for other dephasing interactions, such as
elastic collisions in an atomic vapor, or elastic phonon scattering
in a solid, the so-called pure dephasing process. Pure dephasing
causes the coherent overlap of the upper and lower state wave
functions to decay in time, while not affecting the state
populations. For example, the pure dephasing rate can be small and
ignored for resonant excitation of a single QD at low temperature (6
K) and power density \cite{Gammon96}. While at elevated temperature,
however, experiments \cite{Fan98, Bayer02} reveal a pure dephasing
contribution that dominates excitonic dephasing. Our results are
directly applicable to experimental data presented in Refs.
\cite{Relthmaier04, Yoshie04, Peter05}, for which no theoretical
predictions were previously available.

In this paper, we derive the normal-mode oscillations and spectra of
the photon emitted both in the side direction and in the forward
beam in the cavity-QED strong-coupling regime. We first review the
derivation of the normal mode oscillations and emission spectra
obtained in the Weisskopf-Wigner approximation (WWA)
\cite{Weisskopf30} in Secs. \ref{WWA} and \ref{NMO and ES}. Then we
focus on the influence of the pure dephasing process, treated in the
phase-diffusion model based on a Wiener-Levy process
\cite{Wodkiewicz791}, on the normal-mode oscillations and emission
spectra in Sec. \ref{Dephasing}. We also extend our recently
presented results for single-photon quantum efficiency (QE)
\cite{Cui05} to include the effects of pure dephasing of the atomic
dipole. We find that the depths of the normal-mode oscillations are
reduced, emission spectra are broadened and the QE is decreased in
the presence of pure dephasing.

\section{\label{WWA} Probability-amplitude method in the Weisskopf-Wigner approximation}

\begin{figure}[t]
\includegraphics[width=0.42\textwidth]{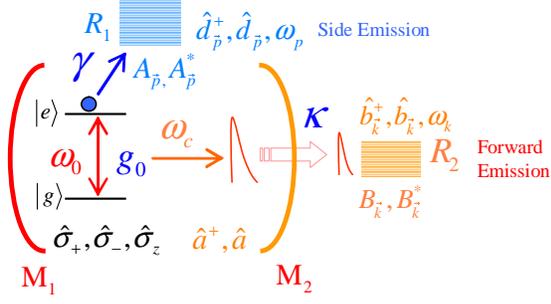}
\caption{(Color online) Schematic description of a lossy two-level
emitter interacting with a single mode in a leaky optical cavity.
$g_0$ is the coupling constant between the emitter and the cavity
field. $A_{\vec p},\,A_{\vec{p}}^{\ast}$ and $B_{\vec k},\,B_{\vec
k}^{\ast}$ are the coupling constants between the emitter, a single
photon and their respective reservoir fields $(R_1,\,R_2)$.}
\label{fig1}
\end{figure}

Consider the interaction of a quantized radiation field with a
two-level emitter, an atom or QD, located at an antinode of the
field in an optical microcavity with a length L, as in
Fig.~\ref{fig1}. $\mathrm{M_1}$ is a perfect 100\%-reflecting mirror
and $\mathrm{M_2}$ is a partially transparent one, from which a
sequence of single photons-on-demand emerges.

The interaction Hamiltonian $\hat{H}_I(t)$ in the interaction
picture for this system in the dipole approximation and
rotating-wave approximation is \cite{Scully97}
\begin{equation}
\begin{split}
\hat{H}_I(t) =& \hbar g_0 \left(\hat{\sigma}_{+}\hat{a}e^{i\Delta
t}+\mathrm{H.c.} \right)\\
&+ \hbar\sum_{\vec p} \left(A_{\vec p}^{\ast}\hat{\sigma}_{-}
\hat{d}_{\vec p}^{\dag}e^{i\delta_p t}+\mathrm{H.c.} \right)\\
&+\hbar\sum_{\vec k} \left(B_{\vec k}^{\ast}\hat{a} \hat{b}_{\vec
k}^{\dag}e^{i\delta_k t}+\mathrm{H.c.} \right),
\end{split}
\label{eq1}
\end{equation}
where $\Delta=\omega_0-\omega_c,\,\delta_p=\omega_p-\omega_0,\,
\delta_k=\omega_k-\omega_c$ are the detunings of the emitter-cavity,
emitter-reservoir, and cavity-reservoir. $\hat{a}$ and
$\hat{a}^{\dag}$ are the annihilation and creation operators for the
single cavity mode under consideration, while $\hat{\sigma}_z$ and
$\hat{\sigma}_{\pm}$ are the Pauli operators for the emitter
population inversion, raising, and lowering, respectively. Here we
treat the atomic transition frequency $\omega_0$ as constant. Later
we allow it to fluctuate, to model pure dephasing.

Given that there is only one excitation in the system, the state
vector can be written as
\begin{equation}
\begin{split}
\left|\psi(t)\right\rangle = &
E(t)|e,0\rangle|0\rangle_{R_1}|0\rangle_{R_2}+
C(t)|g,1\rangle|0\rangle_{R_1}|0\rangle_{R_2} \\
&+\sum_{\vec{p}}S_{\vec{p}}(t)|g,0\rangle|1_{\vec{p}}\rangle_{R_1}|0\rangle_{R_2}\\
&+\sum_{\vec{k}}O_{\vec{k}}(t)|g,0\rangle|0\rangle_{R_1}|1_{\vec{k}}\rangle_{R_2}
\end{split},
\label{eq2}
\end{equation}
where $|m,n\rangle~(m=e,\,g;\,n=0,\,1)$ denotes the emitter state
(excited state, ground state) with $n$ photons in the cavity. $|
j_{\vec{p}}\rangle_{R_1}|l_{\vec{k}}\rangle_{R_2}~(j,\,l=0,\,1)$
corresponds to $j$ photons in the $\vec{p}$ mode (other than the
privileged cavity mode) of the emitter reservoir $R_1$ and $l$
photons in a single-mode $(\vec{k})$ traveling wave of the
one-dimensional photon reservoir $R_2$ (output beam). $E(t)$,
$C(t)$, $S_{\vec{p}}(t)$, and $O_{\vec{k}}(t)$ are the slowly
varying probability amplitudes.

The equations of motion for the probability amplitudes are obtained
by substituting $|\psi(t)\rangle$ and $\hat{H}_I(t)$ into the
Schr\"odinger equation and then projecting the resulting equations
onto different states respectively. In the WWA \cite{Weisskopf30,
Scully97}, we obtain
\begin{eqnarray}
\dot{E}(t)&=& -ig_0 e^{i\Delta t}C(t)- \gamma E(t), \nonumber \\
\dot{C}(t)&=& -ig_0 e^{-i\Delta t}E(t)- \kappa C(t) \label{eq3} \\
S_{\vec{p}}(t)&=& -iA_{\vec{p}}^{\ast}\int_0^t {dt'' e^{i\delta_p
t''}E(t'')}, \nonumber \\
O_{\vec{k}}(t)&=& -iB_{\vec{k}}^{\ast}\int_0^t {dt' e^{i\delta_k
t'}C(t')} \label{eq4}
\end{eqnarray}
where $\gamma$ and $\kappa$ are one-half the radiative decay rates
of the atomic population (other than the privileged cavity mode) and
the intracavity field, respectively. Dots indicate time derivatives.
The general solutions to the coupled differential Eqs. (\ref{eq3})
and (\ref{eq4}) are
\begin{widetext}
\begin{eqnarray}
\left(
  \begin{array}{c}
    E(t) \\
    C(t) \\
  \end{array}
\right) &=& e^{-(K/2)t} \left( \begin{array}{cc}
   e^{i\Delta t/2} & 0  \\
   0 & e^{-i\Delta t/2} \\
\end{array} \right) \nonumber \\
 && \times \left[ e^{i\lambda t} \left( \begin{array}{cc}
 {\frac{1}{2} - \frac{i\Gamma + \Delta }{4\lambda}} & {-\frac{g_0}{2\lambda}}
 \\[1mm]
   {-\frac{g_0}{2\lambda}} & {\frac{1}{2}+\frac{i\Gamma + \Delta}{4\lambda}} \\
\end{array} \right) + e^{-i\lambda t} \left( \begin{array}{cc}
   {\frac{1}{2} + \frac{i\Gamma + \Delta}{4\lambda}} & {\frac{g_0}{2\lambda}}
   \\[0.8mm]
   {\frac{g_0}{2\lambda}} & {\frac{1}{2}-\frac{i\Gamma + \Delta}{4\lambda}} \\
\end{array} \right) \right] \left( \begin{array}{cc}
   {E(0)} \\
   {C(0)}
\end{array} \right),
\label{eq5}
\end{eqnarray}
\end{widetext}
where $K \equiv\kappa+\gamma,~\Gamma\equiv\kappa-\gamma$, and
$\lambda =\sqrt{g_0^2 - [(\Gamma - i\Delta )/2]^2}$.

In the strong-coupling regime, defined by $g_0 \gg \kappa,\,\gamma$,
the real part of $\lambda$ is much larger than its imaginary part.
Then $\lambda$ can be approximated as $\lambda \approx g \equiv
\sqrt{g_0^2 + (\Delta/2)^2 - (\Gamma /2)^2}$, which is the
generalized vacuum Rabi frequency. Note that for the case when the
emitter and cavity are exactly at resonance, $\Delta =\omega _0 -
\omega _c = 0$, the complex frequency $\lambda$ is purely real and
equals $\sqrt{g_0^2 - (\Gamma /2)^2}$. The solutions to the
probability amplitudes are then
\begin{widetext}
\begin{equation}
\left(\begin{array}{c}
E(t) \\
C(t) \\
\end{array}\right)=
e^{-(K/2)t} \left(\begin{array}{cc}
e^{i\Delta t/2} & 0 \\
0 & e^{-i\Delta t/2} \\
\end{array} \right)
\left(\begin{array}{cc}
\cos(gt) + \frac{\Gamma - i\Delta}{2g}\sin(gt) & - i\frac{g_0}{g}\sin(gt) \\
-i\frac{g_0}{g}\sin(gt) & \cos(gt)-\frac{\Gamma-i\Delta}{2g}\sin(gt)
\\ \end{array} \right)
\left(\begin{array}{c}
E(0)\\
C(0)\\
\end{array} \right).
\label{eq6}
\end{equation}
\end{widetext}
In this context, we assume the quantum emitter is prepared in an
excited state $E(0)=1,\,C(0)=0$ at time $t_0=0$ (more generally, it
can be prepared in an arbitrary single-quantum state). The solutions
subject to this initial condition are
\begin{eqnarray}
E(t) &=& e^{-[({K}-i\Delta)/2]t} \left[\cos(gt)+ \frac{\Gamma -
i\Delta}{2g} \sin(gt) \right]
\label{eq7} \\
C(t) &=& e^{-[({K} + i\Delta )/2]t} \left[ - \frac{ig_0}{g} \sin(gt)
\right]. \label{eq8}
\end{eqnarray}
$S_{\vec p}(t)$ and $O_{\vec k}(t)$ can be obtained by carrying out
the integrations in Eq. (\ref{eq4}).

\section{\label{NMO and ES} Normal-mode oscillations and emission spectra of SPS in the
cavity-QED strong-coupling regime}

The strong interaction between an excited quantum emitter and a
single cavity mode leads to single-quantum Rabi oscillation
(normal-mode oscillation) in the time domain or a frequency
splitting in the frequency domain, the so-called normal-mode
splitting, which arises from the coherent interaction of two
degenerate systems---the single quantum emitter and the single
cavity mode. In this section, we discuss the normal-mode
oscillations and emission spectra of SPS in the cavity-QED
strong-coupling regime. We first investigate the normal-mode
oscillations by calculating the probabilities of finding the
composite system in different states. Then we define and calculate
the spectra of SPS appropriate for this case in the long-time limit.

\subsection{Normal-mode oscillations}

\begin{figure*}[htbp]
\centering
\begin{minipage}[b]{0.46\textwidth}
\centering
\includegraphics[width=0.95\textwidth]{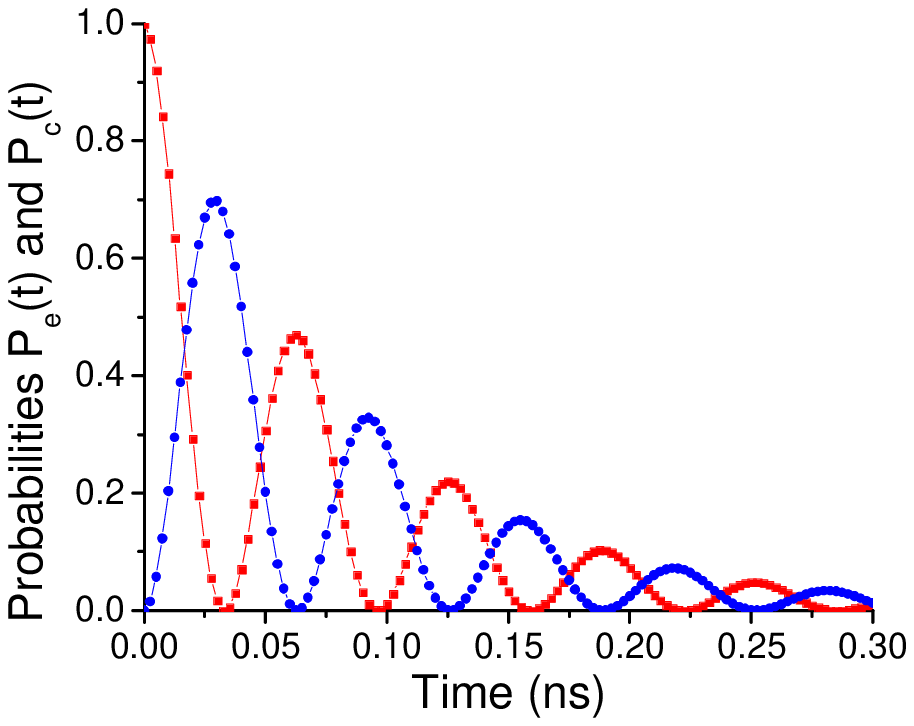}\\
{(a)}
\end{minipage}
\hspace{5mm}
\begin{minipage}[b]{0.45\textwidth}
\centering
\includegraphics[width=0.95\textwidth]{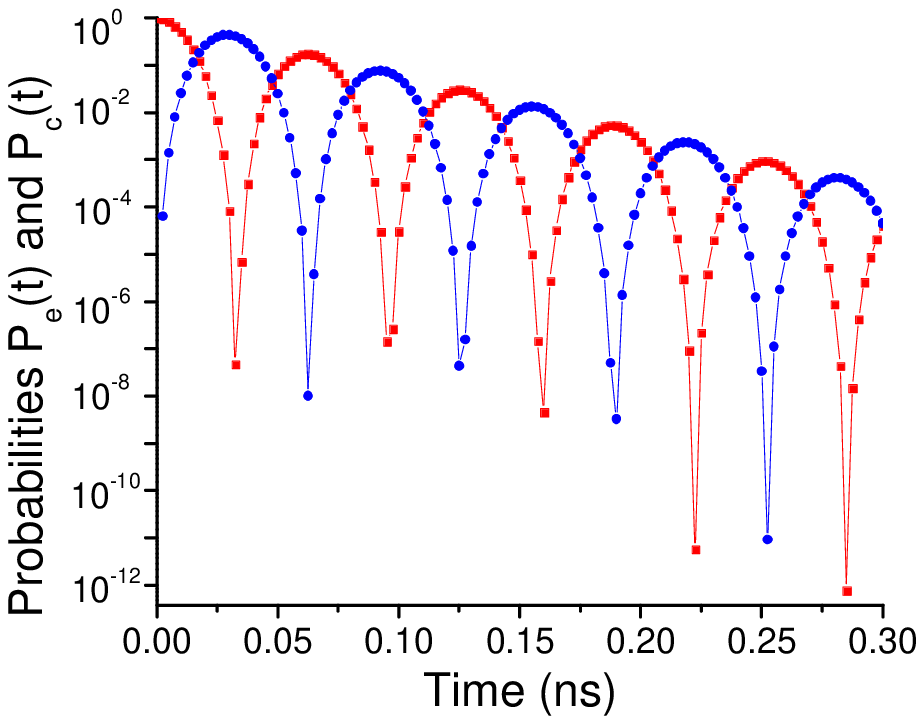} \\
{(b)}
\end{minipage}
\caption{(Color online) Probabilities of finding the system in an
excited atomic state and in a single cavity mode, red square and
blue dot curves respectively, with (a) linear scale and (b)
logarithmic scale, given $(g_0,\,\kappa,\,\gamma )/2\pi =
(8.0,\,1.6,\,0.32)$ GHz.} \label{fig2}
\end{figure*}

The normal-mode oscillations can be viewed either in the
dressed-state picture or in the bare-state picture. Here we look at
them in the bare-state picture where the oscillations can be
relatively easier to find. The probability of finding the system in
the excited atomic state is
\begin{widetext}
\begin{equation}
P_e(t)=\left|E(t)\right|^2=\frac{e^{-Kt}}{2} \left[1 + \frac{\Gamma
^2+\Delta ^2}{4g^2 }+ \left(1 -\frac{\Gamma ^2 + \Delta ^2
}{4g^2}\right)\cos(2gt)+ \frac{\Gamma}{g} \sin(2gt) \right].
\label{eq9}
\end{equation}
\end{widetext}
The probability of finding the system in the single cavity mode is
\begin{eqnarray}
P_c(t)=\left|{C(t)}\right|^2 =\frac{g_0^2}{g^2}e^{-{K}t} \sin^2(gt).
\label{eq10}
\end{eqnarray}
Consider the case when the emitter and cavity are exactly at
resonance, $\Delta=\omega _0 -\omega _c=0$. Figure~\ref{fig2} are
plots of the two probabilities with both linear and logarithmic
scales.
The probabilities oscillate sinusoidally with an exponential decay
envelope. However, they have opposite phases, which indicate the
coherent oscillatory energy exchange between the excited emitter and
the cavity field.

Define the emission probability $P_o(t)$ to be the probability of
finding a single photon in the output mode of the cavity between the
initial time $t_0=0$ and a later time $t$. This equals
\begin{equation}
\begin{split}
P_o&(t)= 2\kappa \int_0^t {dt' \left|C(t')\right|^2 }=\eta _q \\
& \left\{1 - e^{-{K}t} \left[1 + \frac{{K}^2}{2g^2} \sin ^2(gt)+
\frac{{K}}{2g} \sin(2gt) \right] \right\}
\end{split}
\label{eq11}
\end{equation}
where $\eta _q \equiv \left[g_0^2 / (g_0^2+\kappa \gamma)\right]
\left[\kappa/(\kappa+\gamma)\right]$ is the single-photon QE, given
by the single-photon emission probability $P_o(t)$ in the
sufficiently long-time limit $t \gg K^{-1}$, which has been
discussed in our previous publication \cite{Cui05}.

\subsection{Emission spectra}

It may seem strange to talk about the spectrum of a single-mode
field since we normally associate a single mode with a single
frequency. Here we are dealing, however, with what should more
correctly be called a quasimode, a mode defined in a leaky optical
cavity, which therefore has a finite linewidth. For a stationary and
ergodic process, the Wiener-Khintchine theorem \cite{Mandel97}
states that the spectrum is given by the Fourier transform of the
two-time correlation function of the radiated field. In the
strong-coupling regime and for an impulsive excitation of the
system, however, this relation between the correlation function and
spectrum fails because the coherent interaction overwhelms the
relaxations here. There is no time \textit{t} after which the
correlation functions depend only on the time difference. Thus the
dipole correlation and the emitted field correlation cannot be
stationary. We use a generalized definition of the Wiener-Khintchine
spectrum appropriate in this case (Appendix~\ref{Spectra}).

The side emission and forward emission spectra are defined as, in the long-time limit ($t\gg K^{-1}$)
\begin{widetext}
\begin{eqnarray}
S_{SE}({\Omega}') &=& \frac{2\gamma}{\pi} {\rm Re} \left\{
{\int_0^\infty {d\tau e^{i{\Omega}' \tau} \left[ {\int_0^\infty {dt}
E(t + \tau )E^*(t)} \right]} } \right\},
\label{eq12} \\
S_{FE}(\Omega ) &=& \frac{2\kappa}{\pi} {\rm Re} \left\{
{\int_0^\infty {d\tau e^{i\Omega \tau} \left[ {\int_0^\infty {dt}
C(t + \tau )C^*(t)} \right]} } \right\},
\label{eq13}
\end{eqnarray}
\end{widetext}
where ${\Omega}' \equiv \omega - \omega_0$ is the side emission frequency centered at the atomic transition frequency $\omega_0$, and $\Omega \equiv \omega - \omega _c$ is the forward emission frequency centered at the cavity resonance $\omega_c$. Note that ${\Omega}' \equiv \Omega - \Delta$, they are equal when the atom-cavity is at resonance. Substituting the Eqs. (\ref{eq7}) and (\ref{eq8}) into Eqs. (\ref{eq12}) and (\ref{eq13}), we obtain the unnormalized spectra
\begin{eqnarray}
S_{SE}({\Omega}') = \frac{\gamma }{\pi} \left| \frac{\kappa
-i({\Omega}' + \Delta)}{\left({K}/2 - i\Delta /2 - i{\Omega}'\right)^2 +
g^2} \right|^2,
\label{eq14} \\
S_{FE}(\Omega) = \frac{\kappa }{\pi } \left| \frac{-ig_0
}{\left({K}/2 + i\Delta /2 - i\Omega \right)^2 + g^2} \right|^2.
\label{eq15}
\end{eqnarray}

When writing them explicitly in terms of frequency variables $\omega$, $\omega_0$, and $\omega_c$ instead of $\Omega'$, $\Omega$, and $\Delta$, Eqs. (\ref{eq14}) and (\ref{eq15}) change to
\begin{eqnarray*}
S_{SE}(\omega - \omega_0) = \frac{\gamma }{\pi} \left| \frac{\kappa
-i({\omega} - \omega_c)}{\left[{K}/2 - i{(2\omega -\omega_0 - \omega_c)}/2\right]^2 +
g^2} \right|^2,\\
S_{FE}(\omega - \omega_c) = \frac{\kappa }{\pi } \left| \frac{-ig_0
}{\left[K/2 - i{(2\omega -\omega_0 - \omega_c)}/2\right]^2 + g^2} \right|^2.
\end{eqnarray*}
The side emission and forward emission spectra are typically measured independently in experiment. The detuning $\Delta$ in two cases are realized differently. For example, to measure the side emission spectrum, we keep the atomic transition frequency $\omega_0$ fixed while tuning the cavity resonance $\omega_c$. On the other hand, to measure the forward emission spectrum, the cavity resonance $\omega_c$ is fixed and the atomic transition frequency $\omega_0$ is varied.

\begin{figure}[htbp]
\centering
\begin{minipage}[b]{0.49\textwidth}
\centering
\includegraphics[width=0.98\textwidth]{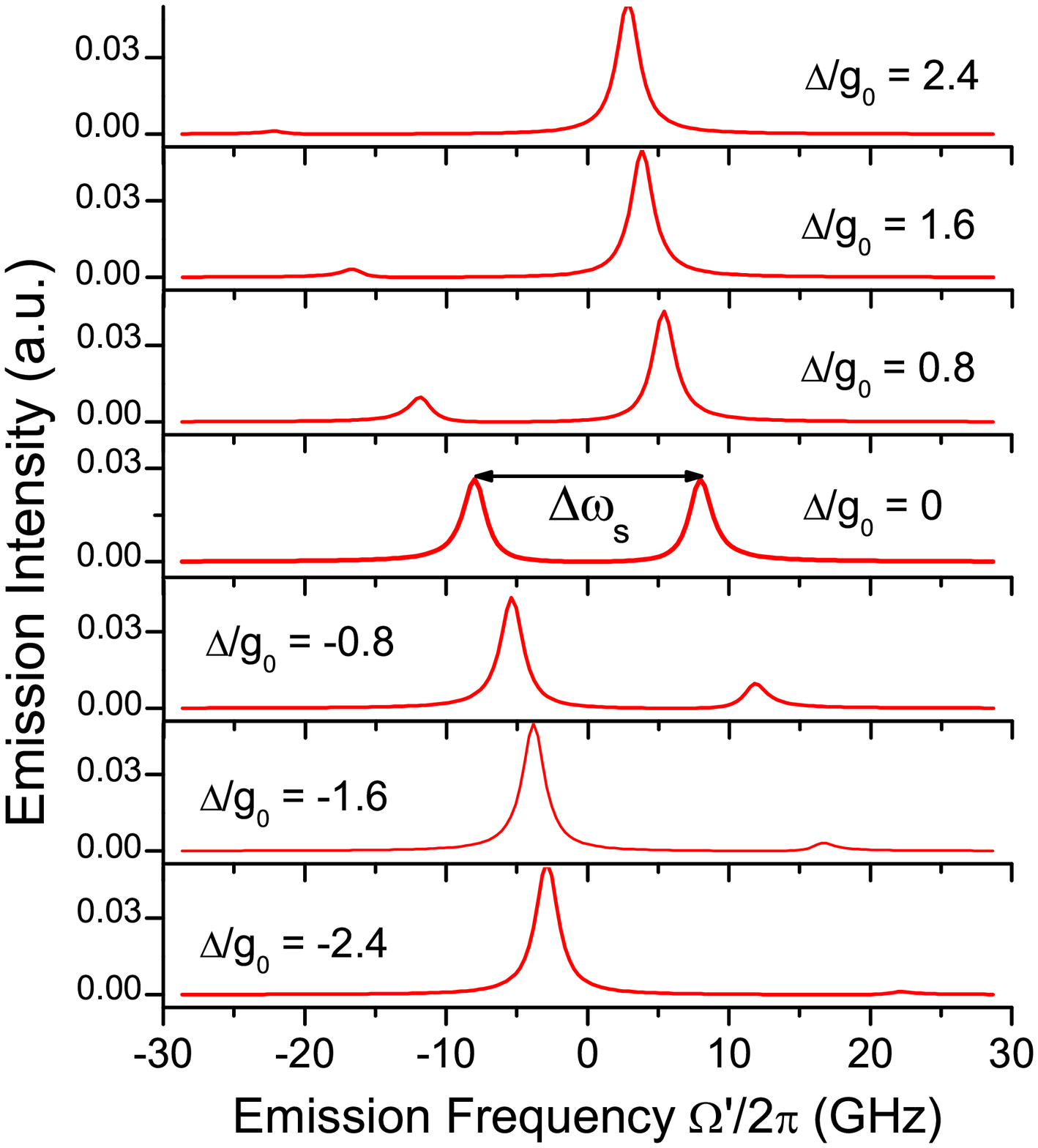} \\
{(a)}
\end{minipage}
\vspace{0.5mm}
\begin{minipage}[b]{0.49\textwidth}
\centering
\includegraphics[width=0.98\textwidth]{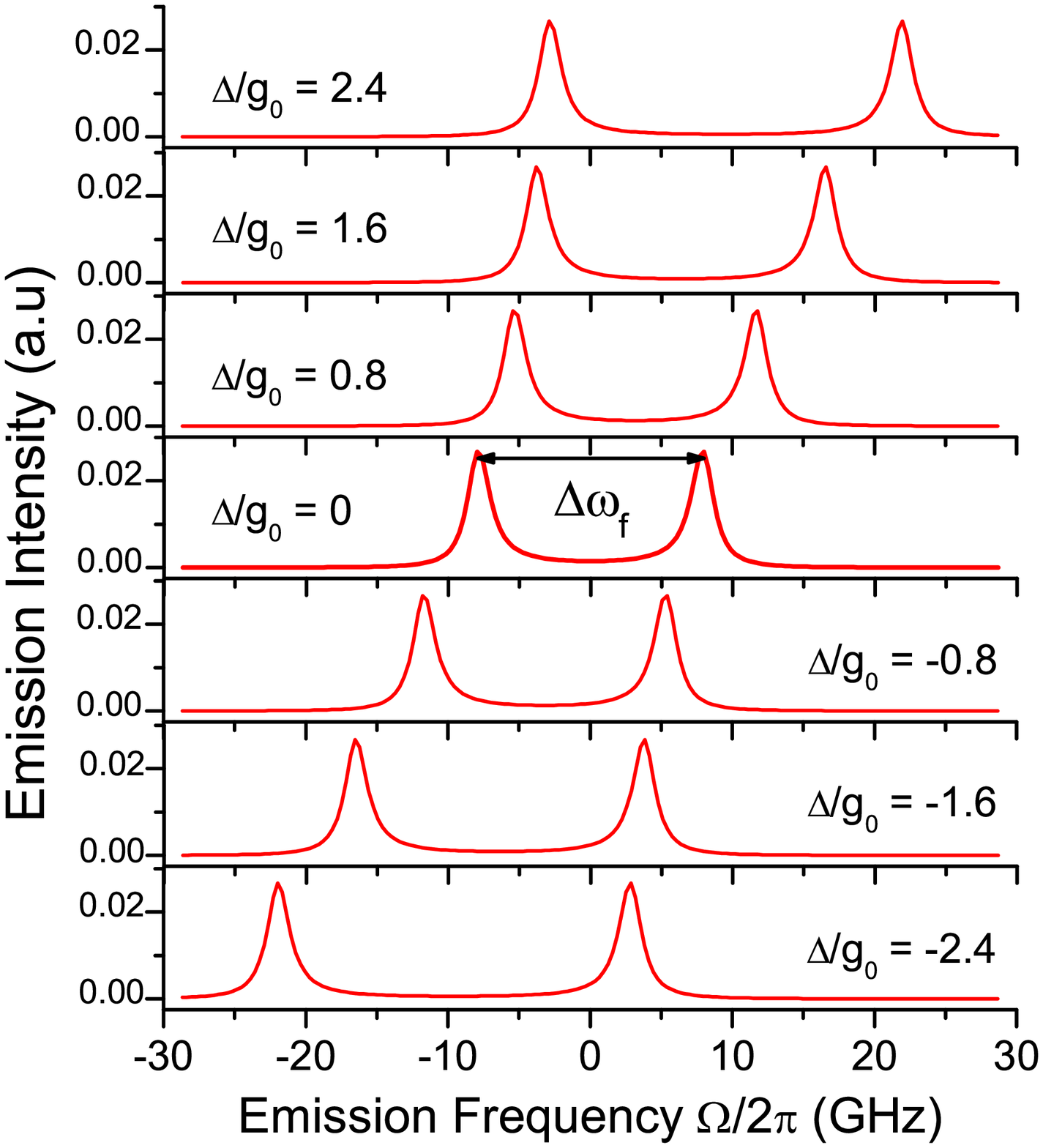} \\
{(b)}
\end{minipage}
\caption{(Color online) Normalized (a) side emission spectra, and
(b) forward emission spectra of SPS for seven different values of
atom-cavity detuning $\Delta/g_0 =
(2.4,\,1.6,\,0.8,\,0,\,-0.8,\,-1.6,\,-2.4)$, given $(g_0
,\,\kappa,\,\gamma )/2\pi = (8.0,\,1.6,\,0.32)$ GHz.}
\label{fig3}
\end{figure}

The side emission spectrum here is the same as the spontaneous
emission spectrum calculated elsewhere \cite{Carmichael89}. The
forward emission spectrum is what we expect to measure by an ideal
detection system at the output of the cavity in the forward
direction, and has not been presented previously, to our knowledge.
For zero atom-cavity detuning $\Delta=0$, where the atom and cavity
resonances are degenerate, both the side emission and forward
emission spectra show the normal-mode splittings, which however, are
different. The splittings are $\Delta\omega _s = 2\sqrt{[g_0^4 +
2g_0^2 \kappa(\kappa + \gamma)]^{1/2} - \kappa^2}$ for the side
emission, and $\Delta \omega _f = 2\sqrt{g_0^2 -(\kappa ^2 + \gamma
^2)/2}$ for the forward emission, as shown by the thicker red curves
in Figs.~\ref{fig3}(a) and~\ref{fig3}(b) respectively. Both are
different from the generalized Rabi splitting $2g$.

Beyond the energy-splitting difference at zero atom-cavity detuning,
it is also illuminating to investigate the dependence of the energy
eigenvalue structure on the atom-cavity detuning. Shown in
Fig.~\ref{fig3} are plots of the spectra, in the strong-coupling
regime, for seven different values of atom-cavity detuning $\Delta$.
As $|\Delta|$ increases, the vacuum Rabi splitting also increases
for both the side emission and forward emission spectra. At the same
time, for the side emission spectra, the cavitylike peak features
stronger emission and the atomlike peak grows smaller. While for the
forward emission spectra, however, both peaks show the same emission
intensity.

\section{\label{Dephasing} Influence of pure dephasing on the normal-mode oscillations
and emission spectra}

Generally speaking, pure dephasing means the decay of the dipole
coherence without change in the populations of the system. Any real
transition to other states leads to population decay. Thus the pure
dephasing is caused by virtual processes which start from a relevant
state and, through some excursion in the intermediate states, return
to the same initial state. These virtual processes give rise to the
temporal fluctuations of phases of the wave functions, which
consequently lead to pure dephasing.

\subsection{Phase-diffusion model of pure dephasing}

\begin{figure}[htbp]
\centering
\includegraphics[width=0.31\textwidth]{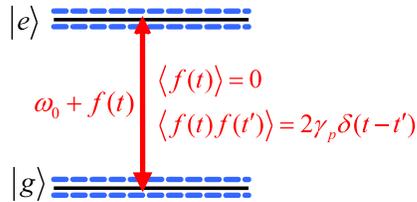}
\caption{(Color online) Schematic diagram of pure dephasing process
in the phase-diffusion model.}
\label{fig4}
\end{figure}

The effects of pure dephasing can be calculated numerically, based
on the Green function formalism by considering the microscopic
details of various virtual processes \cite{Takagahara03}. Instead,
for simplicity we treat this problem analytically in the
phase-diffusion model where the incoherence due to elastic
collisions or elastic phonon scattering is described by a stochastic
model of random frequency modulation, as shown in Fig.~\ref{fig4},
replacing the atomic transition frequency or the phase of the wave
function by an instantaneous one
\begin{equation}
\begin{split}
\omega _0 &\to \omega _0(t) = \omega _0 + f(t) \quad {\rm or} \\
\omega _0 t &\to \int_0^t{dt'\omega _0(t')} = \omega _0 t+
\varphi(t),
\end{split}
\label{eq16}
\end{equation}
where $f(t)$ is the instantaneous deviation of the transition
frequency due to the elastic collisions or scattering process and
$\varphi(t) \equiv \int_0^t{f(t')dt'}$ is the instantaneous
stochastic phase of the wave function.

We assume the phase of the wave function is a Wiener-Levy process
\cite{Wodkiewicz791}. In this phase-diffusion model $f(t) \equiv
\dot \varphi (t)$ is a random, stationary, Gaussian variable with
the mean value and the meansquare correlation given by
\begin{equation}
\left\langle{f(t)}\right\rangle =0, \,\left\langle{f(t)f(t')}
\right\rangle = 2\gamma _p\, \delta(t - t').
\label{eq17}
\end{equation}
The angular brackets indicate a statistical average over the random
variables of the stochastic process. The Markovian nature of the
process is reflected by the presence of the delta function $\delta(t
- t')$. The Gaussian property is introduced such that all higher
correlation functions can be obtained from the second-order
correlation function by permutations and multiplications
\cite{Mandel97}. $2\gamma_p$ is the pure dephasing rate.

Taking into account the pure dephasing modeled by the stochastic
process, the net change for Eq.~(\ref{eq3}) is that the phase term
$e^{i\Delta t}$ should be replaced, such that
\begin{eqnarray}
\dot E(t) &=& -ig_0 e^{i[\Delta t + \varphi(t)]} C(t) - \gamma E(t)
\label{eq18} \\
\dot C(t) &=& -ig_0 e^{- i[\Delta t + \varphi(t)]}E(t)-\kappa C(t).
\label{eq19}
\end{eqnarray}
We note that the above equations with stochastic random variables
are examples of a multiplicative stochastic process, studied
intensively by Refs. \cite{Kubo62, Fox72, Wodkiewicz792}. We solve
these equations exactly, using the method developed by Wodkiewicz
\cite{Wodkiewicz792} for a multiplicative stochastic process
described by the following general vector equation
\begin{equation}
\frac{d}{dt} \vec v(t) =\left[M_0 + if(t)M_1 \right]\vec v(t),
\label{eq20}
\end{equation}
where $\vec v(t)$ is an \textit{n}-dimensional vector, $M_0$ and
$M_1$ are arbitrary $n \times n$ matrices, in general complex and
time independent, and $f(t)$ is the random variable of the
stochastic process described by Eq.~(\ref{eq17}). The equations of
the type (\ref{eq20}) can be solved for the quantum expectation
value of $\vec v(t)$ exactly \cite{Wodkiewicz792}. For a Wiener-Levy
process, the stochastic average of the equation satisfies the
following differential equation
\begin{equation}
\frac{d}{dt} \left\langle{\vec v(t)} \right\rangle = \left[M_0 -
\gamma _p M_1^2 \right] \left\langle{\vec v(t)} \right\rangle.
\label{eq21}
\end{equation}
The solution to Eq.~(\ref{eq21}) can be written in the
Laplace-transform form
\begin{equation}
\left\langle{\vec v(t)} \right\rangle = \int_C {\frac{dz}{2\pi i}
\exp{(zt)} N^{-1}(z) \left\langle{\vec v(0)} \right\rangle},
\label{eq22}
\end{equation}
where the matrix $N^{-1}(z)$ is the inverse to $N(z)$, which itself
is given by the formula $N(z)=z\mathbb{I}-(M_0-\gamma _p M_1^2)$.

In Eq.~(\ref{eq22}), the contour of integration \textit{C} lies
parallel to the imaginary axis in the complex \textit{z} plane, to
the right of all singularities of the integrands. In order to find
the time behavior of $\langle{\vec v(t)}\rangle$, we have to invert
the matrix $N(z)$, whose determinant plays an essential role because
the roots of its secular equation are the poles of the integration
in Eq.~(\ref{eq22}).

\subsection{Normal-mode oscillations and quantum efficiency in the
presence of pure dephasing}

We proceed to solve the stochastic Eqs. (\ref{eq18}) and
(\ref{eq19}). For simplicity, we consider the case when the emitter
and the cavity are in resonance, $\Delta=0$. The detuning can always
be put back without difficulty. First, by making the substitutions
$E(t)=\tilde E(t)e^{-\gamma t}$ and $C(t)=\tilde C(t)e^{ -\kappa t}$
for convenience, we obtain the following simpler equations
\begin{eqnarray}
\dot{\tilde{E}}(t) &=& -ig_0 e^{-\Gamma t}e^{i\varphi(t)}\tilde C(t)
\label{eq23} \\
\dot{\tilde{C}}(t) &=& -ig_0 e^{\Gamma t}e^{-i\varphi(t)}\tilde
E(t).
\label{eq24}
\end{eqnarray}
Then by defining variable $Y(t)=e^{-\Gamma t+i\varphi(t)} \tilde
C(t)$ so that the differential equation for $\tilde E(t)$ does not
explicitly depend on the random variable $\varphi(t)$, we obtain a
matrix equation of the type (\ref{eq20}) for a multiplicative
stochastic process, with
\begin{equation}
\vec v(t) = \left( {\begin{array}{cc}
   \tilde E(t) \\
   Y(t) \\
\end{array}} \right), M_0 = \left( {\begin{array}{cc}
   0 & -ig_0 \\
   -ig_0 & {-\Gamma} \\
\end{array}} \right), M_1 = \left({\begin{array}{cc}
   0 & 0 \\
   0 & 1 \\
\end{array}} \right) \\ \label{eq25}
\end{equation}
and a statistically independent initial condition $\langle{\vec
v(0)}\rangle^T =(\tilde E(0),Y(0))$.

The inverse matrix to the matrix $N(z)$ is
\begin{equation}
N^{-1}(z)=\frac{1}{\det[N(z)]} \left({\begin{array}{cc}
   {z + \Gamma + \gamma _p} & {-ig_0}  \\
   {-ig_0} & z  \\
\end{array}} \right).
\label{eq26}
\end{equation}
Plugging Eq.~(\ref{eq26}) back into Eq.~(\ref{eq22}), using the
Laplace transform technique and choosing properly the contour of
integration \textit{C}, we obtain the quantum expectation value of
the probability amplitude $\langle \tilde E(t) \rangle$,
\begin{widetext}
\begin{eqnarray}
 \left\langle{\tilde E(t)} \right\rangle &=& \int_C {\frac{dz}{2\pi i}e^{zt}
 \frac{(z+\Gamma+\gamma_p)\tilde E(0) - ig_0 Y(0)}{(z - z_1 )(z - z_2 )}} \nonumber \\
 &=& e^{-(\Gamma+\gamma _p)t/2} \left\{ \left[ \cos(g_1 t)+ \frac{\Gamma + \gamma _p}
 {2g_1} \sin(g_1 t) \right] \tilde E(0) - \left[\frac{ig_0}{g_1} \sin(g_1 t)
 \right] Y(0) \right\}, \label{eq27}
\end{eqnarray}
\end{widetext}
where $g_1 \equiv \sqrt{g_0^2-(\Gamma +\gamma _p)^2 /4}$.

Similarly, if we define $X(t)=e^{\Gamma t-i\varphi(t)}\tilde E(t)$,
while keeping $\tilde{C}(t)$ unchanged, we obtain
\begin{widetext}
\begin{eqnarray}
\left\langle{\tilde C(t)}\right\rangle = e^{(\Gamma - \gamma _p)t/2}
\left\{ \left[ \cos(g_2 t)- \frac{\Gamma -\gamma _p}{2g_2} \sin(g_2
t) \right] \tilde C(0)- \left[ \frac{ig_0}{g_2} \sin(g_2 t) \right]
X(0) \right\}, \label{eq28}
\end{eqnarray}
\end{widetext}
where $g_2 \equiv \sqrt{g_0^2-(\Gamma-\gamma _p)^2/4}$. Taking into
account the definitions of $\tilde E(t)$ and $\tilde C(t)$, as well
as the fact that $X(0) = \tilde E(0)= E(0)$ and $Y(0) = \tilde C(0)
= C(0)$, we transform back to $E(t)$ and $C(t)$,
\begin{widetext}
\begin{eqnarray}
\left({\begin{array}{cc}
   {\left\langle{E(t)}\right\rangle } \\
   {\left\langle{C(t)}\right\rangle } \\
\end{array}} \right) = e^{-({\rm{K}}+ \gamma_p)t/2} \left( {\begin{array}{cc}
{\cos(g_1 t)+\frac{\Gamma+\gamma_p}{2g_1} \sin(g_1 t)} & {-\frac{ig_0}{g_1} \sin(g_1 t)}\\
{-\frac{ig_0}{g_2} \sin(g_2 t)} & {\cos(g_2 t)-\frac{\Gamma-\gamma _p}{2g_2} \sin(g_2 t)}\\
\end{array}} \right)
\left({\begin{array}{cc}
   E(0) \\
   C(0) \\
\end{array}} \right). \label{eq29}
\end{eqnarray}
\end{widetext}
The generalized Rabi frequencies for $\langle{E(t)}\rangle$ and
$\langle{C(t)}\rangle$ now are different from each other, as
compared to Eq.~(\ref{eq6}), where they were the same for both
$\langle{E(t)}\rangle$ and $\langle{C(t)}\rangle$. This implies the
destroying of coherence between the two eigenstates of the system,
due to the pure dephasing process. We will show this
phase-destroying effect on the normal-mode oscillations explicitly
later in this section.

We are more interested in finding the influence of the pure
dephasing on the probabilities $|{C(t)}|^2$ and $|E(t)|^2$, or
$I(t)\equiv |{\tilde C(t)}|^2$ and $J(t)\equiv |\tilde E(t)|^2$,
because they give the normal-mode oscillations and are what one
measures in experiment. In order to find the equations of motion for
them, we have to introduce two other one-time functions $H(t) \equiv
\tilde E(t)\tilde C^*(t)$ and $H^*(t) \equiv \tilde E^*(t)\tilde
C(t)$. The equations of motion for these functions are

\begin{widetext}
\begin{eqnarray}
\frac{d}{dt} \left({\begin{array}{c}
   H(t)  \\
   H^*(t)\\
   I(t)  \\
   J(t)  \\
\end{array}} \right) = \left( {\begin{array}{cccc}
   0 & 0 & {-ig_0 e^{-\Gamma t+ i\varphi(t)}} & {ig_0 e^{\Gamma t + i\varphi(t)}} \\
   0 & 0 & {ig_0 e^{-\Gamma t- i\varphi(t)}} & {-ig_0 e^{\Gamma t - i\varphi(t)} } \\
   {-ig_0 e^{\Gamma t-i\varphi(t)}} & {ig_0 e^{\Gamma t + i\varphi(t)}} & 0 & 0  \\
   {ig_0 e^{-\Gamma t-i\varphi(t)}} & {-ig_0 e^{-\Gamma t+i\varphi(t)}} & 0 & 0
\end{array}} \right) \left( {\begin{array}{c}
   H(t)  \\
   H^*(t)\\
   I(t)  \\
   J(t)  \\
\end{array}} \right). \label{eq30}
\end{eqnarray}
\end{widetext}
We still solve these equations assuming the quantum emitter is
prepared in an excited state $E(0)=1,\,C(0)=0$ at time $t_0=0$. We
solve these one-time functions one by one as we did above for
solving $\langle{E(t)}\rangle$ and $\langle{C(t)}\rangle$. For
example, to find the solution to $\langle{I(t)}\rangle $, defining
$U_I(t)= e^{\Gamma t - i\varphi(t)}H(t)$, $U_I^* = e^{\Gamma t +
i\varphi(t)}H^*(t)$, $Z_I(t)= e^{2\Gamma t}J(t)$ and keeping $I(t)$
unchanged, we obtain a matrix equation as the standard vector form
of Eq.~(\ref{eq20}), with
\begin{widetext}
\begin{eqnarray}
\vec v_I(t) = \left( {\begin{array}{c}
   U_I (t)  \\
   U_I^*(t) \\
   I(t)   \\
   Z_I(t)  \\
\end{array}} \right),\,
M_0 =\left( { \begin{array}{cccc}
   \Gamma  & 0 & {-ig_0} & {ig_0}  \\
   0 & \Gamma  & {ig_0} & {-ig_0}  \\
   {-ig_0} & {ig_0} & 0 & 0  \\
   {ig_0} & {-ig_0} & 0 & 2\Gamma \\
\end{array}} \right),\,
M_1  = \left( {\begin{array}{ccccc}
   -1 & 0 & 0 & 0  \\
   0 & 1 & 0 & 0  \\
   0 & 0 & 0 & 0  \\
   0 & 0 & 0 & 0  \\
\end{array}} \right)
\label{eq31}
\end{eqnarray}
\end{widetext}
and a statistically independent initial condition $\langle{\vec
v_I(0)}\rangle^T=(0,\,0,\,0,\,1)$, where we have used the initial
conditions at $t_0=0$, as well as the definitions of $\tilde E(t)$,
$\tilde C(t)$ and $\vec v_I(t)$.

In the cavity-QED strong-coupling regime, $(4g_0^2 - \Gamma^2) \gg
\Gamma^2,\gamma _p^2$, the solution for $\langle{I(t)}\rangle$ is
found to be well approximated by (see Appendix~\ref{I})
\begin{widetext}
\begin{eqnarray}
\left\langle{I(t)}\right\rangle = \frac{g_0^2}{2g^2} e^{\left[\Gamma
- \gamma _p(1 + \varepsilon)/2 \right]t} \left[ e^{\gamma _p(1 +
3\varepsilon)t/2} -
 \frac{\gamma _p}{4g} \sin(2gt) - \cos(2gt)\right], \label{eq32}
\end{eqnarray}
\end{widetext}
where $\varepsilon \equiv(\Gamma /2g)^2$, and $g \equiv \sqrt{g_0^2
-(\Gamma/2)^2}$ is the generalized Rabi frequency, as defined
before. Treating $\gamma _p /g$ as a perturbation parameter, we kept
the order to $O(\gamma _p /g)$ in the coefficients and the order to
$O(\gamma _p \varepsilon /g)$ in the exponential arguments.

Similarly, after some tedious algebra, we find the time evolutions
of $\langle{J(t)}\rangle$ and $\langle{H(t)}\rangle$ are (see
Appendixes \ref{J} and \ref{H}),
\begin{widetext}
\begin{eqnarray}
\left\langle{J(t)}\right\rangle &=& \frac{g_0^2}{2g^2} e^{ -[\Gamma
+ \gamma _p (1 + \varepsilon)/2 ]t } \left\{e^{\gamma_p
(1+3\varepsilon)t/2}- \left[ \frac{\gamma _p}{4g}- \frac{g(\Gamma
-\gamma_p /2)}{g_0^2} \right] \sin(2gt) - \left(1 -
\frac{2g^2}{g_0^2} \right) \cos(2gt) \right\} \label{eq33} \\
\left\langle{H(t)}\right\rangle &=& \frac{ig_0}{2g}e^{-\gamma_p
(3+\varepsilon)t/2} \left[ \frac{3\Gamma}{2g}e^{\gamma_p
(1-7\varepsilon)t/2} - \frac{\Gamma - \gamma _p
}{g}e^{-\gamma_p(1-9\varepsilon)t/2}- \frac{\Gamma + 2\gamma _p}{2g}
\cos(2gt) + \sin(2gt) \right]. \label{eq34}
\end{eqnarray}
\end{widetext}
Finally, the quantum expectation value of the complex conjugate of
$H(t)$ is just the complex conjugate of its quantum expectation
value $\langle{H^*(t)}\rangle = \langle{H(t)}\rangle ^*$. Using the
definitions of $\tilde E(t)$ and $\tilde C(t)$, we can easily find
the solutions for $\langle{|E(t)|^2}\rangle = e^{-2\gamma t} \langle
{J(t)}\rangle$ and $\langle{|C(t)|^2}\rangle = e^{-2\kappa t}
\langle{I(t)}\rangle$.

\begin{figure*}[htbp]
\centering
\begin{minipage}[b]{0.44\textwidth}
\centering
\includegraphics[width=0.95\textwidth]{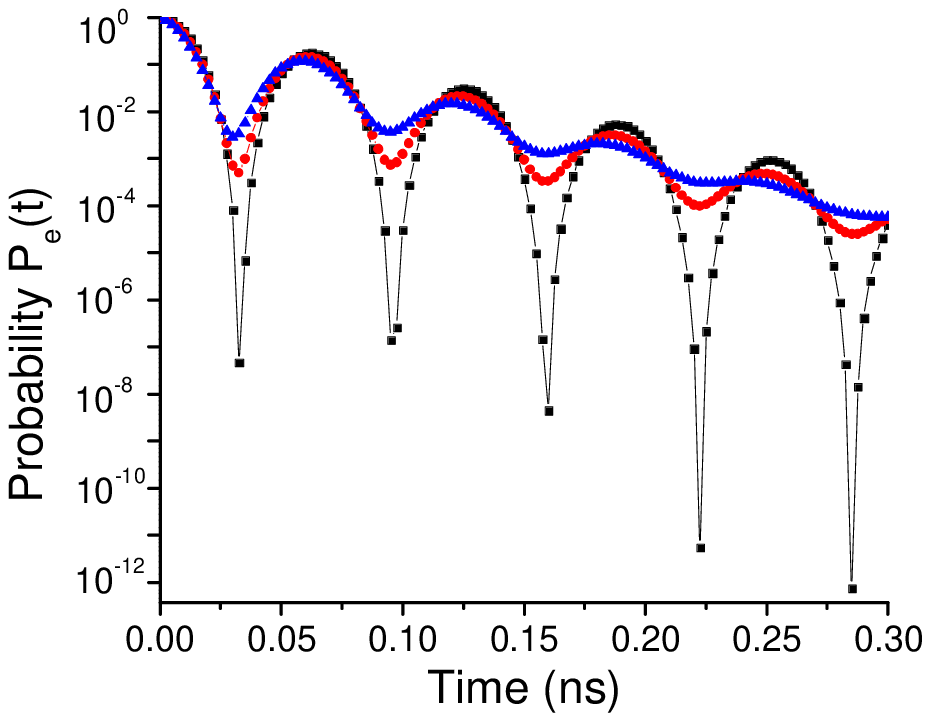}\\
{(a)}
\end{minipage}
\begin{minipage}[b]{0.44\textwidth}
\centering
\includegraphics[width=0.95\textwidth]{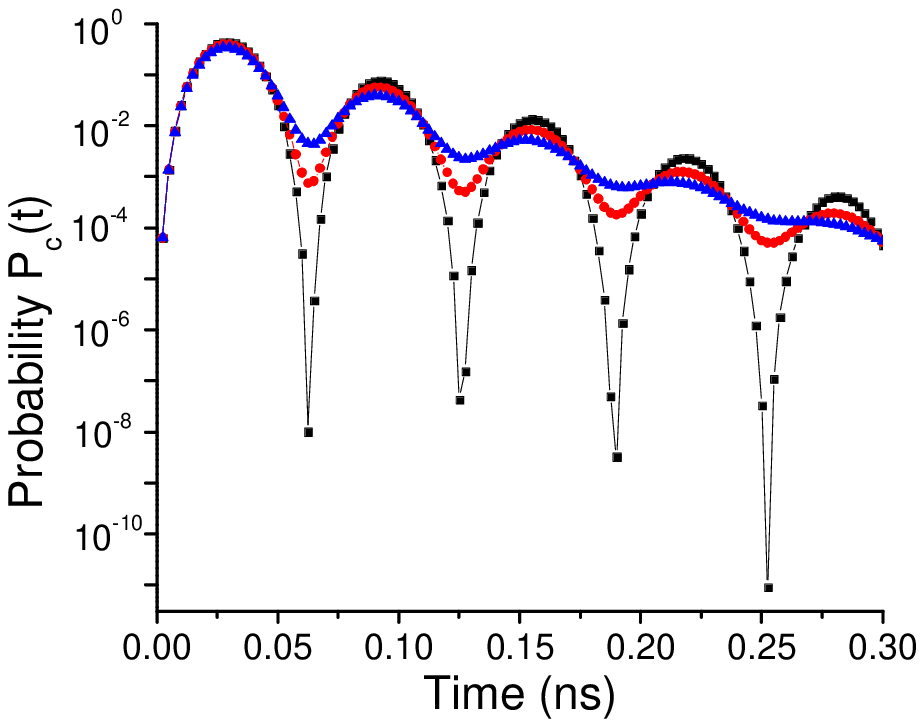} \\
{(b)}
\end{minipage}
\caption{(Color online) Probabilities of finding the system (a) in
the excited atomic state and (b) in the cavity mode, with
logarithmic scale for three different pure dephasing rates $\gamma
_p /2\pi =(0,\,1.0,\,2.5)$ GHz (black square, red dot and blue
triangle respectively), given $(g_0 ,\,\kappa,\,\gamma )/2\pi =
(8.0,\,1.6,\,0.32)$ GHz.}
\label{fig5}
\end{figure*}

Therefore, the probability of finding the system in the excited
atomic state, including the pure dephasing, is
\begin{widetext}
\begin{eqnarray}
\left\langle{P_e(t)}\right\rangle =
\left\langle{|E(t)|^2}\right\rangle = \frac{g_0^2 }{2g^2} e^{-[ {\rm
K}+\gamma_p(1+\varepsilon)/2 ]t } \left\{ e^{\gamma_p
(1+3\varepsilon)t/2}-\left[\frac{\gamma _p }{4g}-\frac{g
(\Gamma-\gamma_p /2)}{g_0^2}\right] \sin(2gt) - \left(
1-\frac{2g^2}{g_0^2}\right) \cos(2gt)\right\}. \label{eq35}
\end{eqnarray}
\end{widetext}
And the probability of finding the system in the single cavity mode
with pure dephasing process is
\begin{multline}
\left\langle{P_c(t)}\right\rangle = \left\langle{|C(t)|^2
}\right\rangle = \frac{g_0^2}{2g^2} e^{ -[{K} + \gamma _p (1 +
\varepsilon)/2 ]t } \\
\times \left[e^{\gamma _p(1+3\varepsilon)t/2} - \frac{\gamma _p}{4g}
\sin(2gt) - \cos(2gt) \right]. \label{eq36}
\end{multline}
Shown in Figs.~\ref{fig5}(a) and~\ref{fig5}(b) are three plots of
each probability in the presence of pure dephasing,

The modulation depths of the red-dot and blue-triangle curves, with
pure dephasing rates $\gamma _p /2\pi = (1.0,\,2.5)$ GHz, are
reduced, as compared with the black-square curves where there is no
pure dephasing. The normal-mode oscillation frequency seems
unaffected because we solved for the probabilities only up to first
order in $\gamma_p /g$. In fact, it will change slightly from $2g$
to $2g(1-\gamma_p^2/32g^2)$ if we approximate to second order in
$\gamma_p /g$. The normal-mode oscillations are smeared in the
presence of the pure dephasing process.

\begin{figure*}[htbp]
\centering
\begin{minipage}[b]{0.46\textwidth}
\centering
\includegraphics[width=0.94\textwidth]{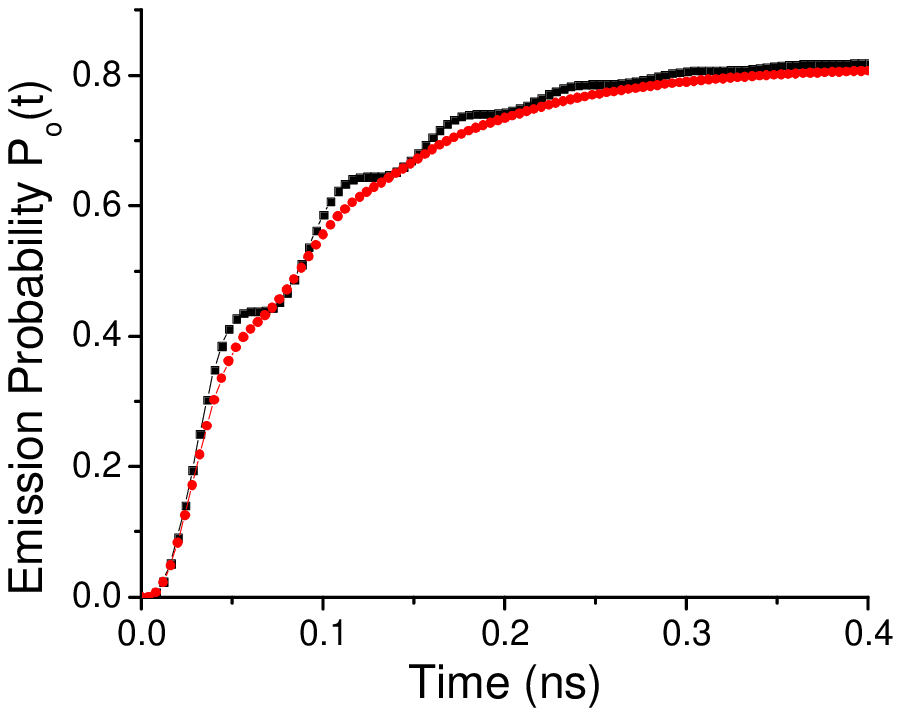}\\
{(a)}
\end{minipage}
\begin{minipage}[b]{0.46\textwidth}
\centering
\includegraphics[width=0.94\textwidth]{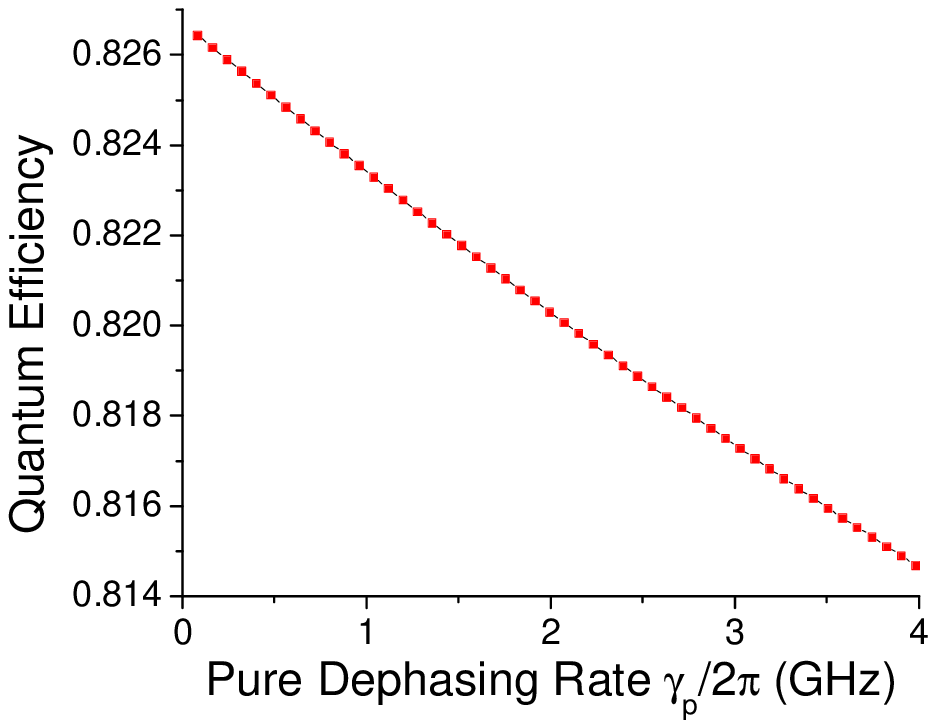} \\
{(b)}
\end{minipage}
\caption{(Color online) (a) Emission probability without and with
pure dephasing rate $\gamma _p /2\pi=4$ GHZ (black square and red
dot curves respectively). (b) The QE $\eta_q$ as a function of the
pure dephasing rate. Other parameters the same as in
Fig.~\ref{fig2}.} \label{fig6}
\end{figure*}

Consequently, the emission probability of a single photon into the
forward beam and the QE with the pure dephasing process, as defined
before, are
\begin{widetext}
\begin{eqnarray}
\left\langle{P_o(t)}\right\rangle &=& 2\kappa \int_0^t{dt'
\left\langle|C(t')|^2 \right\rangle}=\frac{\kappa g_0^2}{g^2}
\left\{ \frac{1-e^{-({K}-\gamma _p \varepsilon )t}}{{K} - \gamma_p
\varepsilon} - \frac{ {K} + \gamma _p (1 + \varepsilon /2)}{\left[
{K} + \gamma _p (1 + \varepsilon )/2 \right]^2 +
(2g)^2}\right\}+ \nonumber \\
&& \frac{\kappa g_0^2}{g^2} \frac{e^{-\left[{K} + \gamma _p (1 +
\varepsilon)/2 \right]t}}{\left[{K} + \gamma _p
(1 + \varepsilon )/2 \right]^2 + (2g)^2} \nonumber \\
&& \left\{ \left[{K}+\gamma _p(1 + \varepsilon)/2 \right]\left[
\frac{\gamma _p}{4g} \sin(2gt)+ \cos(2gt) \right] + 2g
\left[\frac{\gamma _p}{4g} \cos(2gt)- \sin(2gt)\right]
\right\} \label{eq37} \\[2mm]
\eta_q(\gamma _p) &\equiv& \langle{P_o(t \to \infty)}\rangle =
\frac{g_0^2}{g^2} \frac{\kappa}{{K} - \gamma _p \varepsilon} \left\{
1 - \frac{({K}- \gamma _p \varepsilon ) [{K} + \gamma _p (1 +
\varepsilon /2)]}{\left[{K} + \gamma _p (1 + \varepsilon
)/2\right]^2 + (2g)^2} \right\}. \label{eq38}
\end{eqnarray}
\end{widetext}
They reduce to our earlier results, given by and below
Eq.~(\ref{eq11}), in the limit $\gamma _p \to 0$.
Figure~\ref{fig6}(a) are plots of the emission probabilities with
and without pure dephasing, and Fig.~\ref{fig6}(b) is the QE
$\eta_q$ as a function of the pure dephasing rate.
The emission probability is also smeared for a pure dephasing rate
$\gamma_p/2\pi = 4$ GHz compared with no pure dephasing. The QE
decreases only about 1\% as the dephasing rate increases from 0 to 4
GHz.

\subsection{Two-time correlation functions and the emission spectra
in the presence of pure dephasing}

In order to calculate the emission spectra, we need to find the
two-time correlation functions because the emission spectra in the
long-time limit are proportional to the Fourier transform of their
convolutions. The two-time correlation functions are defined as
follows:
\begin{eqnarray}
Q(t,t')&\equiv&\tilde E(t)\tilde E^*(t'),\, R(t,t') \equiv \tilde
C(t)\tilde E^*(t') \label{eq39} \\
F(t,t')&\equiv& \tilde E(t)\tilde C^*(t'),\, G(t,t') \equiv \tilde
C(t)\tilde C^*(t'). \label{eq40}
\end{eqnarray}
Of these $Q(t,t')$ and $G(t,t')$ are required for calculating the
emission spectra.

The quantum regression theorem \cite{Lax63, Carmichael99}, which
provides a framework to calculate two-time correlation functions,
states that the equations of motion for two-time correlation
functions $Q(t,t')$ and $R(t,t')$ and $F(t,t')$ and $G(t,t')$ with
respect to variable $t$ obey the same equations of motion as those
for $\tilde E(t)$ and $\tilde C(t)$, respectively,
\begin{eqnarray}
\partial _t Q(t,t') &=& -ig_0 e^{ - \Gamma t} e^{i\varphi (t)}
R(t,t'), \label{eq41} \\
\partial _t R(t,t') &=& -ig_0 e^{\Gamma t} e^{ - i\varphi (t)}
Q(t,t'), \label{eq42} \\
\partial _t F(t,t') &=& -ig_0 e^{ - \Gamma t} e^{i\varphi (t)}
G(t,t'), \label{eq43} \\
\partial _t G(t,t') &=& -ig_0 e^{\Gamma t} e^{ - i\varphi (t)}
F(t,t'), \label{eq44}
\end{eqnarray}
but now with initial conditions
\begin{eqnarray}
Q(t=t',t') &=& |{\tilde E(t')}|^2  \equiv
J(t'), \nonumber \\
R(t=t',t') &=& \tilde C(t')\tilde E^*(t') \equiv H^*(t'), \label{eq45} \\
F(t =t',t')&=& \tilde E(t')\tilde C^*(t') \equiv H(t'), \nonumber
\\
G(t = t',t') &=& |{\tilde C(t')}|^2  \equiv I(t'), \label{eq46}
\end{eqnarray}
which are already solved and given explicitly by Eqs. (\ref{eq32})
to (\ref{eq34}).

We now specialize to the case $t \ge t'$ and define $t \equiv t' +
\tau$. We are particularly interested in the expectation values of
$\langle{Q(t'+\tau,t')}\rangle$ and $\langle{G(t'+\tau,t')}\rangle$
as pointed out before. The solutions for their expectation values,
according to the quantum regression theorem, have the same forms as
the solutions for the one-time averages of $\langle{\tilde
E(t)}\rangle$ and $\langle{\tilde C(t)}\rangle$ in Eqs. (\ref{eq27})
and (\ref{eq28}), with the initial conditions given above:
\begin{widetext}
\begin{eqnarray}
\left\langle{Q(t'+\tau,t')}\right\rangle &=& e^{-(\Gamma + \gamma
_p)\tau /2} \left\{ \left[ \cos(g_1 \tau) + \frac{\Gamma + \gamma _p
}{2g_1} \sin(g_1 \tau) \right] \left\langle{J(t')}\right\rangle -
\left[ \frac{ig_0}{g_1} \sin(g_1
\tau)\right] \left\langle{H^*(t')}\right\rangle \right\}, \label{eq47} \\
\left\langle G(t'+ \tau,t')\right\rangle &=& e^{(\Gamma + \gamma _p
)\tau /2} \left\{ \left[ \cos(g_2 \tau) - \frac{\Gamma - \gamma _p
}{2g_2} \sin(g_2 \tau) \right] \left\langle{I(t')}\right\rangle -
\left[ \frac{ig_0}{g_2} \sin(g_2 \tau) \right]
\left\langle{H(t')}\right\rangle \right\}. \label{eq48}
\end{eqnarray}
\end{widetext}

By substituting $\tilde E(t')= E(t')e^{\gamma t'}$, $\tilde C(t')=
C(t')e^{\kappa t'}$ and the initial conditions into Eqs.
(\ref{eq47}) and (\ref{eq48}), we obtain the explicit solutions for
$\langle {E(t'+ \tau )E^*(t')}\rangle$ and $\langle{C(t'+ \tau)C^ *
(t')}\rangle$. The side emission and forward emission spectra are
then
\begin{widetext}
\begin{eqnarray}
 S_{SE} (\Omega ) &=& \frac{{2\gamma }}{\pi }{\mathop{\rm Re}\nolimits}
 \left\{ {\int_0^\infty  {d\tau e^{i\Omega \tau } \left[ {\int_0^\infty  {dt'}
 \left\langle {E(t' + \tau )E^* (t')} \right\rangle } \right]} } \right\} =
 \frac{{g_0^2 }}{{g^2 }} {\mathop{\rm Re}\nolimits} \left\{ {\frac{{\gamma /\pi }}
 {{\left[ {\left( {{K} + \gamma _p } \right)/2 - i\Omega } \right]^2  + g_1^2 }}}
 \right\} \nonumber \\
 && \left\{ {\frac{{ - 3\Gamma }}{{2\left( {{K} + \gamma _p  + 4\gamma _p \varepsilon }
 \right)}} - \frac{{2g^2  - \left[ {{K} + \gamma _p \left( {3 + \varepsilon }
 \right)/2} \right]\left( {\Gamma /2 + \gamma _p } \right)}}{{\left[ {{K} + \gamma _p
 \left( {3 + \varepsilon } \right)/2} \right]^2  + \left( {2g} \right)^2 }} +
 \frac{{\Gamma  - \gamma _p }}{{{K} + 2\gamma _p  - 4\gamma _p \varepsilon }}}
 \right\} + \nonumber \\
 && \frac{{g_0^2 }}{{g^2 }} {\mathop{\rm Re}\nolimits} \left\{ {\frac{{(\gamma /\pi )
 \left( {\kappa  + \gamma _p  - i\Omega } \right)}}{{\left[ {\left( {{K} + \gamma _p }
 \right)/2 - i\Omega } \right]^2  + g_1^2 }}} \right\} \left\{ {\frac{1}{{{K} -
 \gamma _p \varepsilon }} - \frac{{{K} - 4g^2 \kappa /g_0^2  + \gamma _p
 [1 + (1 - 2g^2 /g_0^2 )\varepsilon /2]}}{{\left[ {{K} + \gamma _p
 \left( {1 + \varepsilon } \right)/2} \right]^2  + \left( {2g} \right)^2 }}}
 \right\},
 \label{eq49} \\[2mm]
 S_{FE}(\Omega ) &=& \frac{2\kappa}{\pi}{\rm Re}
 \left\{ {\int_0^\infty  {d\tau e^{i\Omega \tau } \left[ {\int_0^\infty  {dt'}
 \left\langle {C(t' + \tau )C^* (t')} \right\rangle } \right]} } \right\} =
 \frac{{g_0^2 }}{{g^2 }} {\mathop{\rm Re}\nolimits} \left\{ {\frac{{\kappa /\pi }}
 {{\left[ {\left( {{K} + \gamma _p } \right)/2 - i\Omega } \right]^2  + g_2^2 }}}
 \right\} \nonumber \\
 && \left\{ {\frac{{3\Gamma }}{{2\left( {{K} + \gamma _p  + 4\gamma _p \varepsilon }
 \right)}} + \frac{{2g^2  - \left[ {{K} + \gamma _p \left( {3 + \varepsilon }
 \right)/2} \right]\left( {\Gamma /2 + \gamma _p } \right)}}{{\left[ {{K} + \gamma _p
 \left( {3 + \varepsilon } \right)/2} \right]^2  + \left( {2g} \right)^2 }} -
 \frac{{\Gamma  - \gamma _p }}{{{K} + 2\gamma _p  - 4\gamma _p \varepsilon }}}
 \right\} + \nonumber \\
 && \frac{{g_0^2 }}{{g^2 }} {\rm Re} \left\{ {\frac{{(\kappa /\pi )
 \left( {\gamma  + \gamma _p  - i\Omega } \right)}}{{\left[ {\left( {{K} +
 \gamma _p } \right)/2 - i\Omega } \right]^2  + g_2^2 }}} \right\}
 \left\{ {\frac{1}{{{K} - \gamma _p \varepsilon }} - \frac{{{K} + \gamma _p
 \left( {1 + \varepsilon /2} \right)}}{{\left[ {{K} + \gamma _p \left( {1 + \varepsilon }
 \right)/2} \right]^2  + \left( {2g} \right)^2 }}} \right\}. \label{eq50}
\end{eqnarray}
\end{widetext}

\begin{figure*}[htbp]
\centering
\begin{minipage}[b]{0.46\textwidth} \centering
\includegraphics[width=0.95\textwidth]{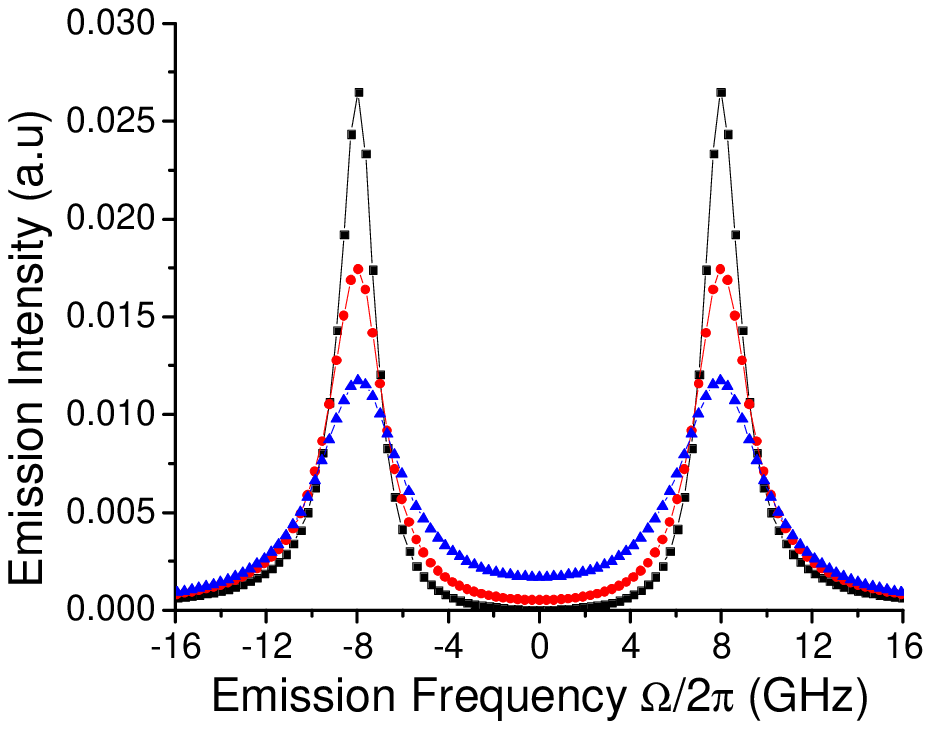}\\
{(a)}
\end{minipage}
\begin{minipage}[b]{0.46\textwidth}
\centering
\includegraphics[width=0.95\textwidth]{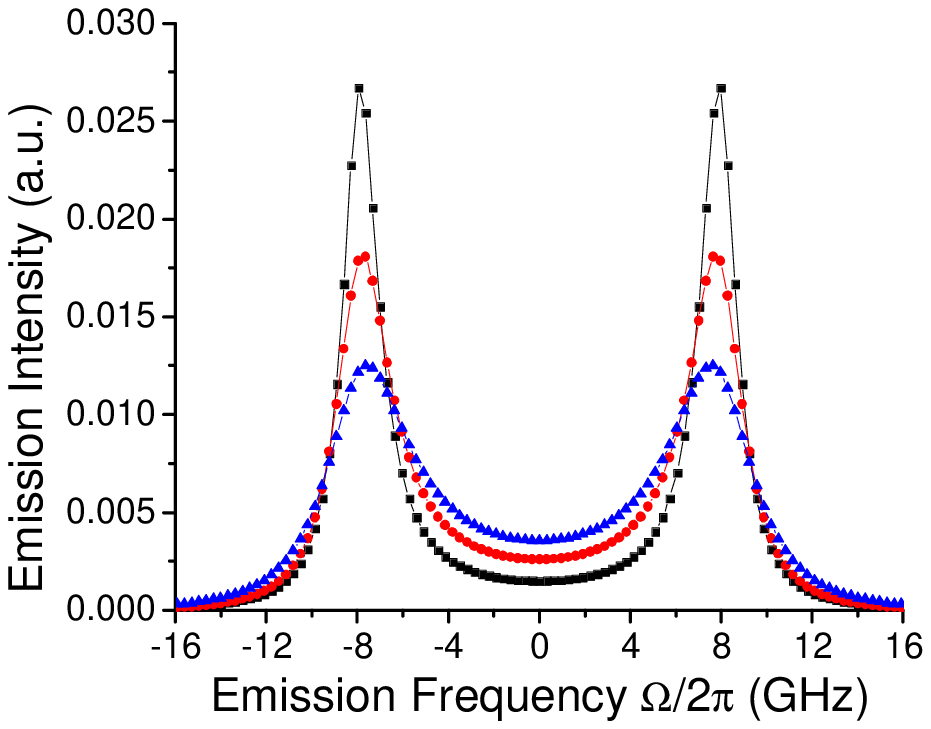} \\
{(b)}
\end{minipage}
\caption{(Color online) Plots of the normalized side emission
spectra (a) and forward emission spectra (b) for three values of
different pure dephasing rate $\gamma _p /2\pi  = (0,\,1.0,\,2.5)$
GHz (black square, red dot and blue triangle respectively), given
$(g_0,\,\kappa,\,\gamma )/2\pi  = (8.0,\,1.6,\,0.32)$ GHz.}
\label{fig7}
\end{figure*}

Shown in Figs.~\ref{fig7}(a) and~\ref{fig7}(b) are the side emission
and forward emission spectra in the long-time limit, for different
dephasing rates $\gamma_p$, while other parameters are the same as
those in Fig.~\ref{fig2}. The effect of pure dephasing is twofold.
The phase fluctuations decrease the peak intensities of the spectra,
and broaden the linewidths of the two peaks and hence smear out the
splittings, which correspond to damping rates of the Rabi
oscillations in the time domain as shown in Fig.~\ref{fig5}. This
effect is further seen to increase with increasing values of pure
dephasing rate $\gamma _p$.

\section{Conclusion}

We derived analytical formulas for the side and forward (useful
cavity output) emission spectra of single-photon sources in the
cavity-QED strong-coupling regime. We also studied the influence of
the pure dephasing process on the emission spectra and the QE, in
the case that the pure dephasing rate is significantly less than the
coherent coupling rate, that is, up to first order in $\gamma_p/g$.
These results should be useful in analyzing photoluminescence
spectra from strongly coupled semiconductor-QD microcavities, where
pure dephasing cannot always be assumed negligible because often
temperature tuning of the QD has to be used to tune through cavity
resonance \cite{Relthmaier04, Yoshie04, Peter05}. One can use this
method, for example, to model the time jitter of solid-state SPS,
where the excited state of the QD or color center in diamond is
often populated by spontaneous phonon emission, by averaging over
nonradiative relaxation time. One may also calculate the two-photon
interference visibility assuming having two independent but
identical SPS and investigate how the pure dephasing processes
affect the indistinguishability of the emitted single photons.

\begin{acknowledgments}
This work was supported by the National Science Foundation Grant No.
ECS-0323141.
\end{acknowledgments}

\appendix

\section{\label{Spectra} Forward emission and side emission spectra}

For a stationary and ergodic process, the Wiener-Khintchine theorem
\cite{Mandel97} states that the spectrum is given by the Fourier
transform of the two-time correlation function of the radiated
field. One can easily generalize the definition of the
Wiener-Khintchine spectrum to that of a nonstationary spectrum
appropriate in this case. We define both the side emission, the
spontaneous emission of the excited emitter into the free-space
other than the cavity (side modes or leak modes), and the forward
emission, the emission of single photons through the cavity mirror
into a single wave-packet, outward-traveling wave. We recognize from
Eq.~(\ref{eq4}) in the text that $O_{\vec k}(t)$ is proportional to
the Fourier transform of the probability amplitude $C(t')$,
\begin{equation}
O_{\vec k}(t) = - iB_{\vec k}^* \int_0^t {dt'e^{i\left({\omega _k -
\omega _c } \right)t'} C(t')}.
\label{a1}
\end{equation}

We define the spectrum as the absolute value squared of the Fourier
transform of the probability amplitude in the long-time limit, which
is proportional to the Fourier transform of the convolution of the
probability amplitude, as will be shown later. Therefore, the forward emission spectrum is given by
\begin{equation}
S_{FE}(\omega - \omega_c) = \mathop {\lim }\limits_{t \to \infty }
D(\omega _c)\left|{O_\omega(t)} \right|^2, \label{a2}
\end{equation}
where we have changed the probability amplitude from $O_{\vec k}
(t)$ to $O_\omega(t)$ by using the density of states for the
one-dimensional photon reservoir $D(\omega _c)= {L/2\pi c}$
\cite{Carmichael99}. Using the solution to the probability amplitude
$C(t)$ and the expression of $O_{\vec k}(t)$ in Eq.~(\ref{a1}), we
can calculate the spectrum
\begin{eqnarray}
S_{FE}(\omega - \omega _c) &=& \mathop {\lim }\limits_{t \to \infty
} D(\omega _c )\left|{B(\omega _c)} \right|^2\int_0^t dt'e^{i\left(
{\omega - \omega _c} \right)t'} \nonumber \\
&& \times C(t') \int_0^t {dt''e^{ - i\left(
{\omega - \omega _c} \right)t''} C^* (t'')}. \label{a3}
\end{eqnarray}
Then using the definition of the decay rate of the intracavity field
$\kappa = \pi D(\omega _c) |B(\omega _c)|^2$ \cite{Carmichael99} and
defining a new variable $\Omega \equiv \omega - \omega _c$, the
forward emission frequency centered at the cavity resonance $\omega_c$, and
$\tau \equiv t' - t''$, we obtain the forward emission spectrum
\begin{multline}
S_{FE}(\Omega ) = \frac{2\kappa}{\pi} \\
\times {\rm Re} \left\{
\int_0^\infty d\tau e^{i\Omega \tau}\left[\int_0^\infty {dt'C(t'+
\tau)C^* (t')} \right] \right\}. \label{a4}
\end{multline}
The normalized forward emission spectrum is
\begin{eqnarray}
s_{FE} = \left( {2\kappa \int_0^\infty  {dt\left| {C(t)} \right|^2
} } \right)^{-1} S_{FE}. \label{a5}
\end{eqnarray}

Similarly the side emission spectrum and the normalized side
emission spectrum, in the long time limit, are given by
\begin{multline}
S_{SE}({\Omega}') = \frac{2\gamma}{\pi} \\
\times {\rm Re} \left\{
{\int_0^\infty {d\tau e^{i{\Omega}' \tau} \left[ {\int_0^\infty {dt}
E(t + \tau )E^*(t)} \right]} } \right\}, \label{a6}
\end{multline}
\begin{equation}
s_{SE} = \left( {2\gamma \int_0^\infty  {dt\left| {E(t)}
\right|^2 } } \right)^{ - 1} S_{SE}. \label{a7}
\end{equation}
where ${\Omega}' \equiv \omega - \omega_0$ is the side emission frequency centered at the atomic transition frequency $\omega_0$.

\section{\label{Roots} The approximate solutions for the expectation values of
$I(t)$, $J(t)$ and $H(t)$}

\subsection{\label{I} Approximate roots of the secular
equation of the matrix $N_I(z)$ and the solution for $\left\langle
{I(t)} \right\rangle $ in the limit $\left( {4g_0^2 - \Gamma ^2 }
\right) \gg \Gamma ^2 ,\gamma _p^2$}

From Eq.~(\ref{eq31}) in the text, define matrix $M \equiv M_0 -
\gamma _p M_1^2$, given explicitly by
\begin{eqnarray}
M = \left( {\begin{array}{cccc}
   {\Gamma - \gamma _p} & 0 & {-ig_0} & {ig_0}  \\
   0 & {\Gamma - \gamma _p} & {ig_0} & {-ig_0}  \\
   {-ig_0} & {ig_0} & 0 & 0  \\
   {ig_0} & {-ig_0} & 0 & {2\Gamma } \\
\end{array}} \right). \label{b1}
\end{eqnarray}
Then the matrix $N_I(z) \equiv z\mathbb{I}-M$ and its determinant
are
\begin{eqnarray}
N_I(z) = \left( {\begin{array}{cccc}
   {z + \gamma _p - \Gamma} & 0 & {ig_0} & {-ig_0}  \\
   0 & {z + \gamma _p -\Gamma} & {-ig_0} & {ig_0}  \\
   {ig_0} & {-ig_0} & z & 0  \\
   {-ig_0} & {ig_0} & 0 & {z - 2\Gamma} \\
\end{array}} \right) \label{b2}
\end{eqnarray}
and
\begin{equation}
\begin{split}
\det&\left[N_I(z)\right] = \left({z + \gamma _p - \Gamma}
\right) \\
& \left[ {z\left( {z - 2\Gamma } \right)\left( {z + \gamma _p -
\Gamma } \right) + 4g_0^2 \left( {z - \Gamma } \right)} \right].
\end{split} \label{b3}
\end{equation}

The secular equation is given by the vanishing of the determinant
Eq.~(\ref{b3}), which reduces to a cubic equation, for $z_1= \Gamma
- \gamma _p$,
\begin{eqnarray}
({z - \Gamma })\left[( {z - \Gamma })^2 + \gamma _p ( {z - \Gamma }
) + 4g_0^2  - \Gamma ^2\right] = \gamma _p \Gamma ^2, \label{b4}
\end{eqnarray}
which is the standard Torrey equation \cite{Torrey49, Allen87}. This
cubic equation can be solved exactly \cite{Weisstein3}, although
only in an implicit form. As Torrey has pointed out, in the special
case of interest, this equation has a relatively simple explicit
solution. We solve it in the strong-coupling regime, $\left( {4g_0^2
- \Gamma ^2 } \right) \gg \Gamma ^2 ,\gamma _p^2$, in which case
there are two kinds of roots. The first of these follows the
assumption that $\left( {z - \Gamma } \right)^2 $ is small compared
with $\left( {4g{}_0^2  - \Gamma ^2 } \right)$, allowing one to
rearrange the cubic equation (\ref{b4}) and solve by iteration
\begin{eqnarray}
z - \Gamma  &=& \frac{{\gamma _p \Gamma ^2 }}{{4g_0^2  - \Gamma ^2
}}\left[ {1 + \frac{{\left( {z - \Gamma } \right)\left( {z - \Gamma
+ \gamma _p } \right)}}{{4g_0^2  - \Gamma ^2 }}} \right]^{ - 1},
\nonumber \\
z_2 &\approx& \Gamma  + \gamma _p \varepsilon  + O\left[ {\left(
{\frac{{\gamma _p }}{g}} \right)^3 } \right], \label{b5}
\end{eqnarray}
where $\varepsilon\equiv(\Gamma/2g)^2$, and note that $g\equiv
\sqrt{g_0^2-(\Gamma/2)^2}$. The second kind of root occurs when
$\left( {z - \Gamma } \right)^2 $ is as large as $\left( {4g{}_0^2 -
\Gamma ^2 } \right)$, but with opposite sign. The cubic equation
(\ref{b4}) can be written as
\begin{eqnarray}
( {z -\Gamma } )^2  + {4g_0^2 - \Gamma ^2}= - \gamma _p ( {z -
\Gamma } ) \left[ {1 - \frac{{\Gamma ^2 }}{{( {z - \Gamma } )^2 }}}
\right]. \label{b6}
\end{eqnarray}
To first order in $\gamma_p$, the factor $(z - \Gamma)^2$ on the
right-hand side, Eq.~(\ref{b6}) can be replaced by $-(4 g_0^2
-\Gamma ^2)$. This gives a quadratic equation for $(z - \Gamma)$,
$(z - \Gamma)^2 + \gamma _p (1 + \varepsilon)(z - \Gamma) + 4g^2 =
0$, whose solutions are the third and fourth roots
\begin{eqnarray}
 z_{3,4}&\approx& \Gamma  - \frac{{\gamma _p }}{2}(1 + \varepsilon )
 \pm i2g\sqrt {1 - (\gamma _p /4g)^2 }  + O\left[ {\left(
 {\frac{{\gamma _p }}{g}} \right)^2 } \right] \nonumber \\
& \approx& \Gamma  - \frac{{\gamma _p }}{2}(1 + \varepsilon ) \pm
i2g + O\left[ {\left( {\frac{{\gamma _p }}{g}} \right)^2 } \right].
\label{b7}
\end{eqnarray}

The inverse matrix to the matrix $N_I (z)$ in Eq.~(\ref{b2}) is
\begin{eqnarray}
N_I^{-1}(z) = \frac{1}{{\det \left[N_I(z)\right]}}\left(
{\begin{array}{cccc}
    \bullet  &  \bullet  &  \bullet  &  \bullet \\
    \bullet  &  \bullet  &  \bullet  &  \bullet \\
   {n_{31}} & {n_{32} } & {n_{33} } & {n_{34} } \\
    \bullet  &  \bullet  &  \bullet  &  \bullet \\
\end{array}} \right) \nonumber
\end{eqnarray}
with
\begin{eqnarray}
 n_{31} &=& - ig_0 \left( {z - 2\Gamma } \right)\left( {z + \gamma _p  - \Gamma } \right) \nonumber \\
 n_{32} &=& ig_0 \left( {z - 2\Gamma } \right)\left( {z + \gamma _p  - \Gamma } \right) \nonumber \\
 n_{33} &=& \left( {z + \gamma _p  - \Gamma } \right)\left[ {\left( {z - 2\Gamma }\nonumber \right)
 \left( {z + \gamma _p  - \Gamma } \right) + 2g_0^2 } \right] \\
 n_{34} &=& 2g_0^2 \left( {z + \gamma _p  - \Gamma } \right) \nonumber
\end{eqnarray}
where we only calculate the elements of the third row of $N_I^{-1}
(z)$ because they are required to calculate $\left\langle{I(t)}
\right\rangle$, which is then
\begin{widetext}
\begin{eqnarray}
 \left\langle {I(t)} \right\rangle  &=& \int_C {\frac{{dz}}{{2\pi i}}e^{zt}
 \frac{{n_{34} }}{{\left( {z - z_1 } \right)\left( {z - z_2 } \right)\left( {z - z_3 }
 \right)\left( {z - z_4 } \right)}}}=\int_C {\frac{{dz}}{{2\pi i}}e^{zt} \frac{{2g_0^2 }}
 {{\left( {z - z_2 } \right)\left( {z - z_3 } \right)\left( {z - z_4 } \right)}}} \nonumber \\
 &\approx& \frac{{g_0^2 }}{{2g^2 }}e^{\left[ {\Gamma  - \gamma _p
 \left( {1 + \varepsilon } \right)/2} \right]t}  \left[ {e^{\gamma _p
 \left( {1 + 3\varepsilon } \right)t/2}  - \frac{{\gamma _p }}{{4g}}\sin (2gt) -
 \cos (2gt)} \right], \label{b8}
\end{eqnarray}
\end{widetext}
where we have used the initial condition that $\langle{\vec v_I
(0)}\rangle^T=(0,\,0,\,0,\,1)$ Treating $\gamma _p /g$ as a
perturbation parameter, we kept the order to $O(\gamma _p /g)$ in
the coefficients and the order to $O(\gamma _p \varepsilon /g)$ in
the exponential arguments.

\subsection{\label{J} Approximate roots of the secular
equation of the matrix $N_J(z)$ and the solution for $\left\langle
{J(t)} \right\rangle $ in the limit $\left( {4g_0^2  - \Gamma ^2 }
\right) \gg \Gamma ^2 ,\gamma _p^2 $}

As it is clear from the definition of $\vec v_I(t)$, we can only
obtain the solution for $\langle{I(t)}\rangle$ in the above
calculation. In order to obtain the solution for
$\langle{J(t)}\rangle$, we have to derive another equation of the
type Eq.~(\ref{eq20}) with the following definitions of the vector
and matrices:
\begin{eqnarray}
\vec v_J (t) &=& \left({\begin{array}{c}
   {U_J (t)}  \\
   {U^ *  _J (t)}  \\
   {W_J (t)}  \\
   {J(t)}  \\
\end{array}} \right),\,
M_1  = \left( {\begin{array}{cccc}
   { - 1} & 0 & 0 & 0  \\
   0 & 1 & 0 & 0  \\
   0 & 0 & 0 & 0  \\
   0 & 0 & 0 & 0  \\
\end{array}} \right), \nonumber \\
M_0 &=& \left( {\begin{array}{cccc}
   { - \Gamma } & 0 & { - ig_0 } & {ig_0 }  \\
   0 & { - \Gamma } & {ig_0 } & { - ig_0 }  \\
   { - ig_0 } & {ig_0 } & { - 2\Gamma } & 0  \\
   {ig_0 } & { - ig_0 } & 0 & 0 \\
\end{array}} \right)
\label{b9}
\end{eqnarray}
and with the initial condition $\langle{\vec v_J (0)}\rangle ^T = (
0,\,0,\,0,\,1)$, where $U_J(t)=e^{-\Gamma t-i\varphi(t)}H(t)$, $U^
*_J(t) = e^{-\Gamma t + i\varphi(t)}H^*(t)$, $W_J (t) = e^{-
2\Gamma t}I(t)$.

The calculation of $N_J(z)$ is almost the same as the calculation in
Appendix \ref{I}. The matrix $M \equiv M_0 - \gamma _p M_1^2$ is
\begin{eqnarray}
M = \left( {\begin{array}{cccc}
   {-\Gamma -\gamma _p} & 0 & {-ig_0} & {ig_0}  \\
   0 & {-\Gamma - \gamma _p} & {ig_0} & {-ig_0}  \\
   {- ig_0} & {ig_0} & {-2\Gamma} & 0  \\
   {ig_0} & {-ig_0} & 0 & 0  \\
\end{array}} \right). \label{b10}
\end{eqnarray}
Then the matrix $N_J(z) \equiv z\mathbb{I} - M$ and its determinant
are, respectively,
\begin{eqnarray}
N_J(z) = \left( {\begin{array}{cccc}
   {z + \gamma _p + \Gamma} & 0 & {ig_0} & {-ig_0}  \\
   0 & {z + \gamma _p + \Gamma} & {-ig_0} & {ig_0}  \\
   {ig_0} & {-ig_0} & {z + 2\Gamma} & 0  \\
   {-ig_0} & {ig_0} & 0 & z  \\
\end{array}} \right) \nonumber
\end{eqnarray}
and
\begin{equation}
\begin{split}
\det&\left[N_J(z)\right] = \left( {z + \gamma _p  + \Gamma } \right)  \\
& \left[ {z\left( {z + 2\Gamma } \right)\left( {z + \gamma _p  +
\Gamma } \right) + 4g_0^2 \left( {z + \Gamma } \right)} \right],
\end{split} \label{b11}
\end{equation}
which is the same as Eq.~(\ref{b3}) provided that we change $\Gamma$
to $-\Gamma$. So the roots of the secular equation of the matrix
$N_J(z)$ are
\begin{equation}
\begin{split}
& z_1 \approx -\Gamma  - \gamma _p , \; z_2 \approx -\Gamma  +
\gamma _p \varepsilon  + O\left[ {\left( {\frac{{\gamma _p }}{g}}
\right)^3
} \right], \\
& z_{3,4} \approx -\Gamma  - \frac{{\gamma _p }}{2}(1 + \varepsilon
) \pm i2g + O\left[ {\left( {\frac{{\gamma _p }}{g}} \right)^2 }
\right].
\end{split} \label{b12}
\end{equation}

The inverse matrix to the matrix $N_J(z)$ is therefore
\begin{eqnarray}
N_J^{-1}(z) = \frac{1}{{\det\left[N_J(z)\right]}}\left(
{\begin{array}{cccc}
    \bullet  &  \bullet  &  \bullet  &  \bullet   \\
    \bullet  &  \bullet  &  \bullet  &  \bullet   \\
    \bullet  &  \bullet  &  \bullet  &  \bullet   \\
   {n_{41}} & {n_{42}} & {n_{43}} & {n_{44}}  \\
\end{array}} \right) \nonumber
\end{eqnarray}
with
\begin{eqnarray}
 n_{41} &=& ig_0 \left( {z + 2\Gamma } \right)\left( {z + \gamma _p  + \Gamma } \right), \nonumber \\
 n_{42} &=& - ig_0 \left( {z + 2\Gamma } \right)\left( {z + \gamma _p  + \Gamma } \right), \nonumber \\
 n_{43} &=& 2g_0^2 \left( {z + \gamma _p  + \Gamma } \right), \nonumber \\
 n_{44} &=& \left( {z + \gamma _p  + \Gamma } \right)\left[ {\left( {z + 2\Gamma }
 \right)\left( {z + \gamma _p  + \Gamma } \right) + 2g_0^2 }
 \right],
 \nonumber
\end{eqnarray}
where we only calculate the elements of the fourth row of
$N_J^{-1}(z)$ because they are required to calculate $\left\langle
{J(t)} \right\rangle $, which is then
\begin{widetext}
\begin{eqnarray}
 \left\langle {J(t)} \right\rangle &=& \int_C {\frac{{dz}}{{2\pi i}}e^{zt} \frac{{n_{44} }}
 {{\left( {z - z_1 } \right)\left( {z - z_2 } \right)\left( {z - z_3 } \right)\left( {z - z_4 }
 \right)}}} = \int_C {\frac{{dz}}{{2\pi i}}e^{zt} \frac{{\left( {z + 2\Gamma }
\right)
 \left( {z + \Gamma  + \gamma _p } \right) + 2g_0^2 }}{{\left( {z - z_2 } \right)
 \left( {z - z_3 } \right)\left( {z - z_4 } \right)}}} \nonumber \\
 &\approx& \frac{{g_0^2 }}{{2g^2 }} e^{- \left[ {\Gamma  + \gamma _p
 \left( {1 + \varepsilon } \right)/2} \right]t}  \left\{ {e^{\gamma _p
 \left( {1 + 3\varepsilon } \right)t/2}  - \left[ {\frac{{\gamma _p }}{{4g}} -
 \frac{{g\left( {\Gamma  - \gamma _p /2} \right)}}{{g_0^2 }}} \right]\sin (2gt) -
 \left( {1 - \frac{{2g^2 }}{{g_0^2 }}} \right)\cos (2gt)} \right\}, \label{b13}
\end{eqnarray}
\end{widetext}
where we have used the initial condition that $\left\langle {\vec
v_J (0)} \right\rangle ^T  = \left(0,\,0,0,1 \right)$ and kept the
order to $O\left( {\gamma _p /g} \right)$ and $O\left( {\Gamma /g}
\right)$ in the coefficients and the order to $O\left( {\gamma _p
\varepsilon /g} \right)$ in the exponential arguments.

\subsection{\label{H} Approximate roots of the secular
equation of the matrix $N_H(z)$ and the solution for $\left\langle
{H(t)}\right\rangle $ in the limit $\left( {4g_0^2  - \Gamma ^2 }
\right) \gg \Gamma ^2 ,\gamma _p^2 $}

In order to obtain the solution for $\left\langle {H(t)}
\right\rangle$, we derive another equation of the type
Eq.~(\ref{eq20}) with the following definitions of the vector and
matrices:
\begin{eqnarray}
\vec v_H (t) &=& \left( {\begin{array}{c}
   {H(t)}  \\
   {U_H (t)}  \\
   {W_H (t)}  \\
   {Z_H (t)}  \\
\end{array}} \right),\,
M_1  = \left( {\begin{array}{cccc}
   0 & 0 & 0 & 0  \\
   0 & 2 & 0 & 0  \\
   0 & 0 & 1 & 0  \\
   0 & 0 & 0 & 1  \\
\end{array}} \right), \nonumber \\
M_0 &=& \left( {\begin{array}{*{20}c}
   0 & 0 & { - ig_0 } & {ig_0 }  \\
   0 & 0 & {ig_0 } & { - ig_0 }  \\
   { - ig_0 } & {ig_0 } & { - \Gamma } & 0  \\
   {ig_0 } & { - ig_0 } & 0 & \Gamma   \\
\end{array}} \right) \nonumber
\end{eqnarray}
and with the initial condition $\left\langle {\vec v_H (0)}
\right\rangle ^T  = \left( {0,0,0,1} \right)$, where $U_H (t) =
e^{i2\varphi(t)}H^*(t)$, $W_H(t) = e^{-\Gamma t + i\varphi(t)}
I(t)$, $Z_H(t)= e^{\Gamma t + i\varphi(t)}J(t)$.

The matrix \textit{M} for the vector $\vec v_H (t)$ is
\begin{eqnarray}
M = M_0 - \gamma _p M_1^2  = \left( {\begin{array}{cccc}
   0 & 0 & { - ig_0 } & {ig_0 }  \\
   0 & { - 4\gamma _p } & {ig_0 } & { - ig_0 }  \\
   { - ig_0 } & {ig_0 } & { - \Gamma  - \gamma _p } & 0  \\
   {ig_0 } & { - ig_0 } & 0 & {\Gamma  - \gamma _p }  \\
\end{array}} \right). \nonumber
\end{eqnarray}
Therefore
\begin{eqnarray}
N_H(z) = \left( {\begin{array}{cccc}
   z & 0 & {ig_0 } & { - ig_0 }  \\
   0 & {z + 4\gamma _p } & { - ig_0 } & {ig_0 }  \\
   {ig_0 } & { - ig_0 } & {z + (\Gamma  + \gamma _p )} & 0  \\
   { - ig_0 } & {ig_0 } & 0 & {z - (\Gamma  - \gamma _p )}  \\
\end{array}} \right) \nonumber
\end{eqnarray}
and the determinant of the matrix $N_H (z)$ is
\begin{multline}
\det[N_H(z)] = z{\left( {z + 4\gamma _p } \right)\left[
{\left( {z + \gamma _p } \right)^2  - \Gamma ^2 } \right]} \\
+ 4g_0^2 \left( {z + \gamma _p } \right)\left( {z + 2\gamma _p }
\right). \label{b14}
\end{multline}
The secular equation is given by setting $\det [N_H(z)] = 0$, which
gives
\begin{multline}
\left( {z + \gamma _p } \right)^2 \left( {z + 2\gamma _p } \right)^2
+ 4g_0^2 \left( {z + \gamma _p } \right)\left( {z +
2\gamma _p } \right)- \\
\Gamma ^2 \left( {z + 2\gamma _p } \right)^2 - 4\gamma _p^2 \left(
{z + \gamma _p } \right)^2  =  - 4\gamma _p^2 \Gamma ^2. \label{b15}
\end{multline}

In the most general case, no simple factorizations occur, and a
quartic equation must be solved. Again the roots are implicit in the
general case \cite{Weisstein4}, but explicit in the strong-coupling
regime. Similarly, there are two kinds of roots in the
strong-coupling regime. The first of these follows from the
assumption that both $\left({z + \gamma _p} \right)^2$ and $\left({z
+ 2\gamma _p} \right)^2 $ are small compared with $\left( {4g_0^2 -
\Gamma ^2} \right)$, in which case it is natural to rearrange
Eq.~(\ref{b15}) into the form
\begin{multline}
\left( {z + \gamma _p } \right)\left( {z + 2\gamma _p } \right)\\
\times \left[ {\left( {z + \gamma _p } \right)\left( {z + 2\gamma _p } \right)
+ \left( {4g_0^2  - \Gamma ^2  - 4\gamma _p^2 } \right)} \right] \approx  -
 4\gamma _p^2 \Gamma ^2, \label{b16}
\end{multline}
where we used the assumptions $({4g_0^2 - \Gamma ^2}) \gg \Gamma ^2
,\,\gamma _p^2$, \\[2mm]
$\left|{\frac{{z + 2\gamma _p }}{{z + \gamma
_p}}} \right| \approx 1 ~\mathrm{and}~ \left|{\frac{{z + \gamma
_p}}{{z + 2\gamma _p }}} \right| \approx 1$. Then
\begin{multline}
\left( {z + \gamma _p } \right)\left( {z + 2\gamma _p } \right)
\approx \\
\frac{- 4\gamma _p^2 \Gamma ^2 }{4g_0^2 - \Gamma ^2 -
4\gamma _p^2 }\left[ {1 + \frac{{\left( {z + \gamma _p }
\right)\left( {z + 2\gamma _p } \right)}}{{4g_0^2  - \Gamma ^2 -
4\gamma _p^2 }}} \right]^{-1}, \label{b17}
\end{multline}
which is solved by iteration. The roots are
\begin{eqnarray}
 z_1 &\approx& - \gamma _p (1 + 4\varepsilon ) + O\left[ {\left( {\frac{{\gamma _p }}{g}}
 \right)^3 } \right] \nonumber \\
 z_2 &\approx& - 2\gamma _p (1 - 2\varepsilon ) + O\left[ {\left(
{\frac{{\gamma _p }}{g}}
 \right)^3 } \right]. \label{b18}
\end{eqnarray}
The second kind of root occurs if $\left( {z + \gamma _p }
\right)\left( {z + 2\gamma _p } \right)$ is as large as $\left(
{4g{}_0^2  - \Gamma ^2 } \right)$, but has the opposite sign. Then
the alternative rearrangement of Eq. (\ref{b15}) is
\begin{multline}
\left( {z + \gamma _p } \right)\left( {z + 2\gamma _p } \right) +
\left( {4g_0^2  - \Gamma ^2  - 4\gamma _p^2 } \right) \\
\approx \gamma _p \Gamma ^2 \left[ {\frac{{z - 2\gamma _p }}{{\left(
{z + \gamma _p } \right)\left( {z + 2\gamma _p } \right)}}} \right].
\label{b19}
\end{multline}
To first order in $\gamma_p$ the factor $\left( {z + \gamma _p }
\right)\left( {z + 2\gamma _p } \right)$ on the right hand side of
Eq.~(\ref{b20}) can be replaced by $-\left( {4g_0^2  - \Gamma ^2 }
\right)$. This gives a simple quadratic equation for \textit{z},
$z^2 + \gamma _p (3 + \varepsilon )z + 4g^2  - 4\gamma _p^2  +
2\gamma _p^2 (1 - \varepsilon ) = 0$, whose solutions are the third
and fourth roots \\[8mm]
\begin{eqnarray}
 z_{3,4} &\approx& - \frac{{\gamma _p }}{2}\left( {3 + \varepsilon }
 \right) \pm i2g\sqrt {1 - (\gamma _p /g)^2 }  + O\left[ {\left(
 {\frac{{\gamma _p }}{g}} \right)^2 } \right] \nonumber \\
 &\approx&  -\frac{{\gamma _p }}{2}\left( {3 + \varepsilon }
 \right) \pm i2g + O\left[ {\left( {\frac{{\gamma _p }}{g}} \right)^2 }
 \right]. \label{b20}
\end{eqnarray}

The inverse matrix to $N_H (z)$ is
\begin{eqnarray}
N_H^{-1}(z) = \frac{1}{{\det[N(z)]}}\left( {\begin{array}{*{20}c}
   {n_{11} } & {n_{12} } & {n_{13} } & {n_{14} }  \\
    \bullet  &  \bullet  &  \bullet  &  \bullet   \\
    \bullet  &  \bullet  &  \bullet  &  \bullet   \\
    \bullet  &  \bullet  &  \bullet  &  \bullet   \\
\end{array}} \right), \nonumber
\end{eqnarray}
with
\begin{eqnarray}
 n_{11}&=& \left( {z + 4\gamma _p } \right)\left[ {\left( {z + \gamma _p }
 \right)^2  - \Gamma ^2 } \right] + 2g_0^2 \left( {z + \gamma _p } \right), \nonumber \\
 n_{12}&=& 0, \nonumber \\
 n_{13}&=& - ig_0 \left( {z + 4\gamma _p } \right)\left( {z + \gamma _p  -
 \Gamma } \right),\nonumber \\
 n_{14}&=& ig_0 \left( {z + 4\gamma _p } \right)\left( {z + \gamma _p  + \Gamma}
 \right). \nonumber
\end{eqnarray}
Therefore $\left\langle {H(t)} \right\rangle$ is given by
\begin{widetext}
\begin{eqnarray}
\left\langle {H(t)} \right\rangle  &=& \int_C {\frac{{dz}}{{2\pi
i}}e^{zt} \frac{{n_{14} }}{{\left( {z - z_1 } \right)\left( {z - z_2
} \right) \left( {z - z_3 } \right)\left( {z - z_4 } \right)}}} =
\int_C {\frac{{dz}}{{2\pi i}}e^{zt} \frac{{ig_0 \left( {z + 4\gamma
_p }\right) \left( {z + \gamma _p  + \Gamma } \right)}}{{\left( {z -
z_1 } \right)
\left( {z - z_2 }\right)\left( {z - z_3 } \right)\left( {z - z_4 } \right)}}}  \nonumber\\
&\approx & \frac{{ig{}_0}}{{2g}}e^{ - \gamma _p (3 + \varepsilon
)t/2}  \left[ {\frac{{3\Gamma }}{{2g}}e^{\gamma _p (1 - 7\varepsilon
)t/2}  - \frac{{\Gamma  - \gamma _p }}{g}e^{ - \gamma _p (1 -
9\varepsilon )t/2}  - \frac{{\Gamma  + 2\gamma _p }}{{2g}}\cos (2gt)
+ \sin (2gt)} \right], \label{b21}
\end{eqnarray}
\end{widetext}
where we have used the initial condition that $\left\langle {\vec
v_H (0)} \right\rangle ^T =\left(0,\,0,\,0,\,1 \right)$ and kept the
order to $O\left( {\gamma _p /g} \right)$ and $O\left( {\Gamma /g}
\right)$ in the coefficients and the order to $O\left( {\gamma _p
\varepsilon /g} \right)$ in the exponential arguments.

\bibliography{Cui_Erratum}

\end{document}